\documentclass[
 reprint,
 floatfix,
 letterpaper,
 superscriptaddress,
 amsmath,amssymb,
 aps,prb
]{revtex4-2}

\usepackage{graphicx}
\usepackage{hyperref}
\hypersetup{colorlinks=true,citecolor=blue,linkcolor=blue,urlcolor=blue}
\graphicspath{{figures/}}

\begin{document}
\raggedbottom

\title{Machine-learned electron densities for self-consistent field calculations in plane-wave Kohn-Sham density functional theory}
\author{Zakaria Elabid}
\affiliation{Center for Advanced Systems Understanding, 02826 G\"orlitz, Germany}
\affiliation{Helmholtz-Zentrum Dresden-Rossendorf, 01328 Dresden, Germany}

\author{Thomas D. K\"uhne}
\affiliation{Center for Advanced Systems Understanding, 02826 G\"orlitz, Germany}
\affiliation{Helmholtz-Zentrum Dresden-Rossendorf, 01328 Dresden, Germany}

\author{Attila Cangi}
\email[Contact author: ]{a.cangi@hzdr.de}
\affiliation{Center for Advanced Systems Understanding, 02826 G\"orlitz, Germany}
\affiliation{Helmholtz-Zentrum Dresden-Rossendorf, 01328 Dresden, Germany}

\date{17 November 2026}

\begin{abstract}
A more accurate starting density can reduce the number of iterations needed to reach self-consistency in a Kohn--Sham density functional theory (DFT) calculation. We test densities from pretrained Materials Learning Algorithms (MALA) models against the usual superposition of atomic densities in Quantum ESPRESSO (QE) calculations for unit cells of aluminum and boron including vacancies. The MALA densities start closer overall to the converged valence densities in every case, but have larger errors near boron vacancies. We use the iteration savings from restarting with the density converged from the atomic guess as a reference. MALA captures 23.08--37.50\% of those savings in pristine cells and up to 16.67\% in cells with vacancies. Both starting guesses give nearly the same final free energies. The iteration savings motivate the development of transferable models that predict starting densities for routine use in Kohn--Sham DFT.
\end{abstract}

\maketitle

\section{Introduction}
\label{sec:introduction}
Kohn--Sham DFT gives the electronic structure and energetics of condensed matter and molecules~\cite{Hohenberg1964,Kohn1965}. At fixed atomic positions, the electron density determines the effective potential. Solving the Kohn--Sham equations in that potential yields a new density, and the cycle repeats until self-consistency. The cost of this self-consistent field (SCF) cycle matters both for individual calculations and for applications that require many such calculations.

The initial density influences the path to self-consistency~\cite{Lehtola2019}. A superposition of atomic densities is inexpensive to construct but omits the redistribution of charge due to bonding and the local environment. A more accurate starting density may reduce the SCF work for the same functional and stopping criterion.

Previous work has shown that learned electronic states can reduce the work of reaching self-consistency. Earlier studies used SchNorb, PhiSNet, and QHNet to supply Hamiltonians or wave functions as starting points~\cite{Schuett2019,Unke2021,Yu2023}. Other studies have predicted density matrices directly with molecular networks and the equivariant Graph2Mat framework~\cite{Hazra2024,Febrer2025}. Analytic refinement has been shown to enforce the electron count and regularize the occupation spectrum before initialization~\cite{YescasRamos2026}. For sequences of configurations, earlier work also combined subsystem contributions to predicted density matrices with extrapolation~\cite{Stishenko2026}. Each representation supplies a starting electronic state from the geometry and requires a corresponding interface to the solver.

Prior studies have demonstrated that the electron density itself can initialize SCF calculations for periodic materials in QE and VASP~\cite{Pope2023,Koker2024,Chen2025ECD}, and for molecules and vacancy-containing silicon supercells in SIESTA~\cite{PantisSimut2026}. Studies of ELECTRA and OrbFlow have tested densities obtained from floating Gaussians and equivariant flow matching, respectively~\cite{Elsborg2025ELECTRA,Liang2026}. Earlier approaches to learning the density include transferable atom-centered and equivariant grid-based models~\cite{Grisafi2019,Jorgensen2022}, pointwise models of the density and local density of states (LDOS)~\cite{Chandrasekaran2019,Ellis2021}, and the residual densities learned by DeepSCF~\cite{Lee2024DeepSCF}. LDOS models have also provided electronic observables for systems larger than those in the training data~\cite{Fiedler2023}.

Transfer across orbital basis sets and exchange-correlation functionals has been demonstrated by using a density predicted in an auxiliary basis to construct the initial Coulomb and exchange-correlation matrices for the usual SCF iteration~\cite{Liu2026}. Previous work has also used densities from pretrained models in the real-space finite-difference code PARSEC.py~\cite{Zhang2026}. Real-space densities provide a route to transfer between solvers, with the predicted valence density still tied to the pseudopotential and representation of the training data.

The utility of an initial guess depends on the solver. Earlier tests of density initialization, comparisons among periodic density, Hamiltonian, and density-matrix initializers, and work on solver-aligned learning have shown that closer agreement with a reference state does not guarantee a shorter path to self-consistency~\cite{Sunshine2023,Chen2026Initializers,Eberhard2026}. The electronic representation also determines which quantities the solver needs. The projector-augmented-wave method requires information beyond the smooth density. In the ChargE3Net benchmarks, the authors used augmentation occupancies from reference calculations~\cite{Koker2024}. A separate preprint examined the effect of supplying augmentation and spin information with the predicted density~\cite{Aertebjerg2026}. The choice of representation can therefore determine whether the target configuration requires converged electronic data for initialization.

In matched plane-wave calculations in QE~\cite{Giannozzi2009,Giannozzi2017}, we compare densities from pretrained MALA models~\cite{Ellis2021,Cangi2025} with QE's superposition of atomic densities, which we label ATOMIC. We compare the first unmixed density outputs and follow the calculations to convergence, using density-only reinjection of the converged ATOMIC density as a reference for the iteration savings.

We test 12 fixed geometries, covering aluminum at 298 K (Al298) and 933 K (Al933), pristine boron at 298 K (B298), and boron systems with vacancies derived from B298.

In Sec.~\ref{sec:methods}, we describe how MALA predicts the density and how we test it in QE. Section~\ref{sec:results} compares the starting densities, first unmixed outputs, and SCF convergence. Section~\ref{sec:conclusions} summarizes the findings and their scope.

Appendix~\ref{app:qe-transfer} describes the transfer of the predicted density to QE. Appendix~\ref{app:parameters} gives the computational settings and additional QE convergence results. Appendix~\ref{app:errors} gives the regional and reciprocal-space error analysis.
Appendix~\ref{app:cp2k-controls} describes separate tests in which we used densities from MALA models trained on plane-wave QE data to start SCF calculations in CP2K, which uses a mixed Gaussian and plane-wave (GPW) representation.

\section{Methods}
\label{sec:methods}

\subsection{Kohn--Sham theory and self-consistency}
\label{sec:ks-background}

The Born--Oppenheimer approximation fixes the ionic positions in the electronic problem. In atomic units, the electron density $n(\mathbf r)$ at position $\mathbf r$ satisfies $\int_{\Omega} n(\mathbf r)\,d\mathbf r=N_e$, where $\Omega$ is the simulation cell and $N_e>0$ is the electron count. For a spin-unpolarized system, the Kohn--Sham Hamiltonian $\hat h_{\mathrm{KS}}[n]$ determines the orbitals and density through~\cite{Hohenberg1964,Kohn1965}
\begin{align}
\hat h_{\mathrm{KS}}[n]\psi_{j\mathbf k} &= \epsilon_{j\mathbf k}\psi_{j\mathbf k},
\label{eq:ks-orbitals}\\
n(\mathbf r) &= \sum_{\mathbf k} w_{\mathbf k}\sum_j f_{j\mathbf k}|\psi_{j\mathbf k}(\mathbf r)|^2\,,
\label{eq:ks-density}
\end{align}
where $\psi_{j\mathbf k}$ is orbital $j$ at wave vector $\mathbf k$ and $\epsilon_{j\mathbf k}$ is its eigenvalue. The Brillouin-zone weights $w_{\mathbf k}$ satisfy $\sum_{\mathbf k}w_{\mathbf k}=1$, and the orbitals are normalized in the cell. The occupation $f_{j\mathbf k}$ includes both spin channels, with $0\leq f_{j\mathbf k}\leq2$. The chemical potential $\mu$ fixes the electron count. A finite electronic temperature $T_e$, distinct from the ionic temperature of the configuration, permits fractional occupations. The variational quantity is the electronic free energy $F=E-T_e S$, where $E$ is the total energy before the entropy correction and $S$ is the electronic entropy~\cite{Mermin1965}.

For local and generalized-gradient exchange-correlation approximations, the Kohn--Sham Hamiltonian is
\begin{equation}
\hat h_{\mathrm{KS}}[n]=-\tfrac12\nabla^2+\hat v_{\mathrm{ext}}+v_{\mathrm H}[n]+v_{\mathrm{xc}}[n].
\label{eq:ks-hamiltonian}
\end{equation}
The kinetic operator is $-\nabla^2/2$, and $\hat v_{\mathrm{ext}}$ is the fixed external ionic potential. The Hartree potential $v_{\mathrm H}$ describes the classical electrostatic interaction of the electron density. The exchange-correlation potential $v_{\mathrm{xc}}=\delta E_{\mathrm{xc}}/\delta n$ is the functional derivative of the exchange-correlation energy $E_{\mathrm{xc}}$ and accounts for many-electron contributions beyond the noninteracting kinetic and Hartree energies. In a generalized-gradient approximation, $E_{\mathrm{xc}}$ depends on the density and its gradient.

The density dependence of $\hat h_{\mathrm{KS}}$ makes Eqs.~\eqref{eq:ks-orbitals} and \eqref{eq:ks-density} nonlinear. At outer SCF iteration $\ell$, diagonalizing the Hamiltonian for the input density $n_{\mathrm{in}}^{(\ell)}$ and occupying its states yields the output density $n_{\mathrm{out}}^{(\ell)}$. Mixing uses this output and, where applicable, earlier iterations to construct the next input. The density residual is
\begin{equation}
R^{(\ell)}(\mathbf r)=n_{\mathrm{out}}^{(\ell)}(\mathbf r)-n_{\mathrm{in}}^{(\ell)}(\mathbf r).
\label{eq:density-residual}
\end{equation}
At self-consistency the input and output densities agree, so the residual vanishes. QE stops at a finite tolerance on $q^{(\ell)}$, the estimated SCF energy error derived from the residual. The corresponding threshold \texttt{conv\_thr} is in Ry per cell (Appendix~\ref{app:parameters}). The unweighted real-space norm of $R^{(\ell)}$ is a distinct quantity.

\subsection{MALA density prediction}
\label{sec:mala}

MALA uses descriptors of the local atomic environment to predict the LDOS, $\widetilde d(\mathbf r,\epsilon)$, as a function of energy $\epsilon$ at each position $\mathbf r$ on a real-space grid~\cite{Ellis2021,Cangi2025}. Integrating over energy with the occupation weights gives the predicted electron density $\widetilde n(\mathbf r)$,
\begin{equation}
\widetilde n(\mathbf r)=\int d\epsilon\,f(\epsilon;\mu,T_e)\,\widetilde d(\mathbf r,\epsilon),
\label{eq:mala-density}
\end{equation}
where $f$ is the Fermi--Dirac distribution at electronic temperature $T_e$ and chemical potential $\mu$. The LDOS $\widetilde d$ includes the factor of 2 for spin degeneracy, consistent with Eq.~\eqref{eq:ks-density}.

We use the previously trained models \texttt{Al\_298K\_0}, \texttt{Al\_933K\_liquid}, and \texttt{boron\_model\_multi\_0}, which are available from the published aluminum and boron data collections~\cite{fiedler2024rodare2725,fiedler2025rodare3746}. Their training LDOS came from QE calculations with norm-conserving (NC) pseudopotentials and $8\times8\times8$ meshes for aluminum and $4\times4\times4$ meshes for boron. The benchmark calculations sample only the Brillouin-zone center $\Gamma$. This change in sampling can contribute to the difference between the prediction and the converged QE density.

We use a single boron model for pristine boron and all nine boron systems with vacancies. For each vacancy geometry, the model predicts the density directly from the remaining atoms. We keep the trained parameters fixed and make no subtraction from the pristine density.

At each grid point, the model maps 91 bispectrum components of the spectral neighbor analysis potential (SNAP) to the LDOS through four hidden layers (Table~\ref{tab:frozen-models}). We evaluate the model with its trained weights and biases and its original input and output scaling.

For a given geometry, we scale the descriptors and evaluate the model on the grid. Integrating the LDOS over space gives the density of states (DOS), from which we determine the chemical potential that yields the prescribed electron count at $T_e$. We then evaluate the energy integral in Eq.~\eqref{eq:mala-density} numerically. For a boron cell with $N_{\rm vac}$ vacancies, we use $N_e=3(144-N_{\rm vac})$ and $T_e=298$~K.

\begin{table*}[tb]
\caption{Architectures and density-generation settings of the previously trained models (Refs.~\onlinecite{fiedler2024rodare2725,fiedler2025rodare3746}), as verified from the model parameters and prediction records. The 91 inputs are bispectrum components, with atomic coordinates excluded as direct inputs. The descriptor cutoff is 4.67637~\AA\ and the band limit is $2j_{\max}=10$ ($j_{\max}=5$). The radial cutoff switching function is disabled at 0 and enabled at 1~\cite{LAMMPSSNAP}. The five affine layers contain $\sum_{i=0}^{4}(w_i+1)w_{i+1}$ weights and biases, where $w_0$ is the input width, $w_1$ through $w_4$ are the hidden widths, and $w_5$ is the output width. The LeakyReLU activation after each layer adds no parameters. The models use feature-wise-standard input scaling and normal output scaling. The LDOS energy grid includes both endpoints and has a spacing of 0.1~eV. Negative predictions are set to zero using \texttt{zero\_out\_negative} before generating the density at the recorded electronic temperature, which can differ from the nominal case label.}
\label{tab:frozen-models}
\small
\setlength{\tabcolsep}{4pt}
\begin{ruledtabular}
\begin{tabular}{lccc}
Parameter & Al298 & Al933 & B298 \\
\hline
Input features & 91 & 91 & 91 \\
Hidden widths & $800,800,800,250$ & $4000,4000,4000,250$ & $800,800,800,250$ \\
LDOS output channels & 250 & 250 & 241 \\
Weights and biases & 1\,618\,200 & 33\,439\,000 & 1\,615\,941 \\
Bispectrum $2j_{\max}$ & 10 & 10 & 10 \\
Cutoff (\AA) & 4.67637 & 4.67637 & 4.67637 \\
\texttt{bispectrum\_switchflag} & 0 & 0 & 1 \\
LDOS interval (eV) & $[-10,14.9]$ & $[-10,14.9]$ & $[-8,16.0]$ \\
LDOS spacing (eV) & 0.1 & 0.1 & 0.1 \\
Recorded $T_e$ (K) & 299.98637 & 931.53663 & 298.0 \\
\end{tabular}
\end{ruledtabular}
\end{table*}

\begin{figure*}[!t]
\centering
\includegraphics[width=\textwidth]{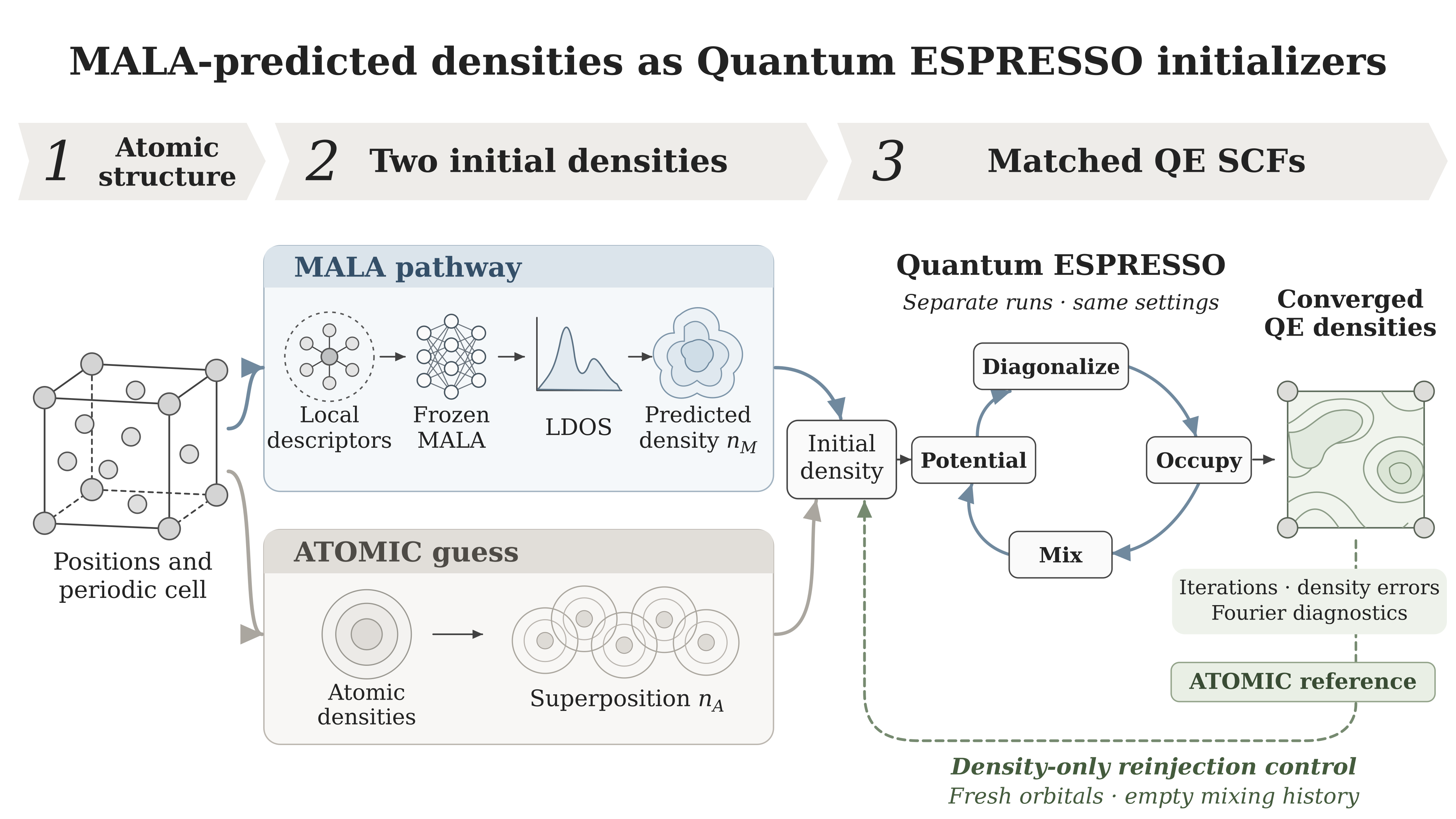}
\caption{MALA and ATOMIC initialization in QE. MALA predicts the starting density $n_{\rm M}$ from the atomic structure and periodic cell. The MALA and ATOMIC guesses initialize independent QE runs with matched settings. The converged ATOMIC valence density $n_{\rm ref}$ serves as the density-error reference and the starting density for the separate reinjection control (dashed arrow). The structures and densities shown are schematic.}
\label{fig:framework}
\end{figure*}

\subsection{Initialization with the predicted density}
\label{sec:initialization}

Figure~\ref{fig:framework} shows how the density prediction enters the QE calculation.

QE expands Kohn--Sham orbitals in plane waves up to $E_{\mathrm{wfc}}$, and the density and local potentials up to $E_\rho$~\cite{Giannozzi2009,Giannozzi2017}. With NC pseudopotentials, Eq.~\eqref{eq:ks-density} gives the valence density directly from occupied orbitals without augmentation. QE uses the Fourier components within the density cutoff,
\begin{equation}
n_{\mathrm{QE}}(\mathbf r)=\sum_{\frac12|\mathbf G|^2\le E_\rho}\widetilde n_{\mathbf G}\,e^{i\mathbf G\cdot\mathbf r},
\label{eq:qe-projection}
\end{equation}
where $\mathbf G$ is a reciprocal-lattice vector with magnitude $G=|\mathbf G|$, and $\widetilde n_{\mathbf G}$ is the corresponding discrete Fourier coefficient of the predicted density. QE constructs the first potential from the projected density $n_{\mathrm{QE}}$. To prepare this input, we use the MALA density and the native-grid metadata. The projection requires no fit to the target orbitals, an atomic-orbital density matrix, or the converged ATOMIC density. Appendix~\ref{app:qe-transfer} gives the transfer procedure.

\subsection{Systems and computational settings}
\label{sec:computational-settings}

The pristine cells include two phases of aluminum, face-centered-cubic Al298 and liquid Al933, and $\alpha$-rhombohedral B298. To vary the local environment in boron, we remove atoms from B298 and keep the cell and the remaining atomic positions fixed. We consider three placements each for one, two, and four vacancies and compare individual geometries. All cells are periodic and have no free surfaces. Table~\ref{tab:benchmark-systems} lists the atom counts, vacancy sites, and valence-electron counts for the neutral cells. Each atom contributes three valence electrons.

\begin{table*}[tb]
\caption{The 12 benchmark configurations. $N_{\rm vac}$ is the number of removed atoms, $N_{\rm at}$ the number remaining, and $N_e$ the valence-electron count. Removed-site indices start at zero in the 144-atom B298 parent cell. A dash denotes a pristine cell. The boron systems with vacancies retain the parent cell and unrelaxed positions of the remaining atoms.}
\label{tab:benchmark-systems}
\small
\setlength{\tabcolsep}{3.5pt}
\begin{ruledtabular}
\begin{tabular}{lrrrl}
Case & $N_{\rm vac}$ & $N_{\rm at}$ & $N_e$ & Removed indices \\
\hline
pristine Al298 & 0 & 256 & 768 & \textemdash \\
pristine Al933 & 0 & 256 & 768 & \textemdash \\
pristine B298 & 0 & 144 & 432 & \textemdash \\
B298, 1 vacancy, set 1 & 1 & 143 & 429 & 89 \\
B298, 1 vacancy, set 2 & 1 & 143 & 429 & 29 \\
B298, 1 vacancy, set 3 & 1 & 143 & 429 & 62 \\
B298, 2 vacancies, set 1 & 2 & 142 & 426 & 5, 106 \\
B298, 2 vacancies, set 2 & 2 & 142 & 426 & 15, 26 \\
B298, 2 vacancies, set 3 & 2 & 142 & 426 & 45, 92 \\
B298, 4 vacancies, set 1 & 4 & 140 & 420 & 5, 34, 88, 95 \\
B298, 4 vacancies, set 2 & 4 & 140 & 420 & 84, 91, 109, 138 \\
B298, 4 vacancies, set 3 & 4 & 140 & 420 & 17, 28, 46, 71 \\
\end{tabular}
\end{ruledtabular}
\end{table*}

We use QE 7.5 with spin-unpolarized occupations. For each element, we keep the exchange-correlation functional and pseudopotential fixed across configurations. Both functionals belong to the Perdew--Burke--Ernzerhof (PBE) family. For aluminum we use the revision for solids, PBEsol~\cite{Perdew2008PBEsol}, with \texttt{Al.sr-pbesol.upf}. For boron we use PBE~\cite{Perdew1996} with \texttt{B\_ncrs\_v041\_pbe.upf}.

The wave-function and density cutoffs set the resolution of the plane-wave calculation. For aluminum, the cutoffs are 100 and 400 Ry on a $200^3$ fast Fourier transform (FFT) grid. For boron, we keep the cutoffs at 80 and 320 Ry on a $108\times108\times135$ grid as vacancies are introduced.

The Fermi--Dirac smearing width is $k_{\rm B}T_e$, where $k_{\rm B}$ is the Boltzmann constant. It sets the electronic temperature independently of the configuration label. The widths are 0.0019, 0.0059, and 0.0018874 Ry for Al298, Al933, and boron, respectively (Table~\ref{tab:qe-parameters}). These values give temperatures close to 298 K for Al298 and boron and 933 K for Al933.

In every pair we use Davidson diagonalization with \texttt{diago\_thr\_init}$=10^{-5}$ Ry and plain modified-Broyden mixing with coefficient 0.1 and history length 8 (Appendix~\ref{app:parameters}). The SCF stopping thresholds are $2.56\times10^{-4}$ Ry/cell for aluminum, $1.44\times10^{-4}$ Ry/cell for pristine boron, and $10^{-6}$ Ry/cell for boron with vacancies. Comparing iteration counts for pristine boron and for boron systems with vacancies therefore includes the effects of both the stopping threshold and the change in geometry.

\subsection{Benchmark design and analysis}
\label{sec:benchmark-protocol}

For each configuration, we use the converged ATOMIC valence density $n_{\rm A,final}$ as the reference $n_{\rm ref}$ for the initial densities, first unmixed outputs, and converged solutions.

\subsubsection{Initial densities}
\label{sec:initial-densities}

We run each configuration to convergence from both starting densities, changing only \texttt{startingpot} within each pair. Both runs use fresh \texttt{atomic+random} wave functions and the same starting Davidson threshold, and begin with an empty mixing history. We compare the densities used to construct the first potential, namely QE's normalized ATOMIC guess and the projected MALA density on the native QE grid (Appendix~\ref{app:qe-transfer}).

\subsubsection{Density-only reinjection controls}
\label{sec:reinjection-protocol}

To measure the SCF work from a converged starting density, we copy the native ATOMIC density file unchanged into a fresh run. QE then rebuilds the orbitals and mixing history using the same Hamiltonian and numerical settings.

\subsubsection{Accuracy of the initial density}
\label{sec:accuracy}

For a valence density $n$ on the native grid, we define the signed electron-count error $\delta N$ relative to the prescribed valence-electron count $N_e$ and the relative $L^2$ density error $\varepsilon_2$ by
\begin{align}
\delta N &= \int_\Omega n(\mathbf r)\,d\mathbf r-N_e,
\label{eq:charge-error}\\
\varepsilon_2 &= \frac{\|n-n_{\mathrm{ref}}\|_2}{\|n_{\mathrm{ref}}\|_2}.
\label{eq:density-l2}
\end{align}
Here $\|u\|_2=(\int_\Omega |u(\mathbf r)|^2\,d\mathbf r)^{1/2}$ is the norm of a field $u$ over the cell, evaluated on the uniform FFT grid. We evaluate these errors for the initial densities and first unmixed outputs. We denote the relative error of the converged MALA density $n_{\rm M,final}$ by $\varepsilon_{2,\rm final}$. To resolve errors around nuclei and removed sites, we restrict both norms to the same region. The root-mean-square error (RMSE) gives the corresponding error in density units. Appendix~\ref{app:errors} defines both spatial partitions and their normalizations.

For nonzero $\mathbf G$, the Hartree kernel weights density errors by $4\pi/G^2$, so errors at long wavelengths can contribute strongly to the potential~\cite{Kerker1981}. We denote by $\bar{v}_{\mathrm H}$ the root-mean-square (RMS) difference between the Hartree potentials of a given density and $n_{\mathrm{ref}}$, with the zero Fourier component removed (Appendix~\ref{app:errors}).

\subsubsection{The first unmixed density response}
\label{sec:first-response}

The first output $n_{\mathrm{out}}^{(1)}$ measures the response to the initial Hamiltonian before density mixing changes the next input. We recover this output by replaying one SCF iteration for each production run. We use these first outputs only to analyze the density response and take the energy trajectories and convergence counts from the production runs (Appendix~\ref{app:qe-transfer}).

\subsubsection{Convergence and final-state agreement}
\label{sec:convergence-analysis}

We count outer SCF iterations from the first diagonalization to the prescribed stopping threshold. The saving is $\Delta N=N_{\rm A}-N_{\rm M}$, where $N_{\rm A}$ and $N_{\rm M}$ are the counts for ATOMIC and MALA. We assess agreement at convergence using $\varepsilon_{2,\rm final}$ and the free energies $F_{\rm M}$ and $F_{\rm A}$, which QE prints for the converged MALA and ATOMIC runs, respectively. For a cell with $N_{\rm at}$ atoms, the absolute free-energy difference per atom is
\begin{equation}
\Delta e=\frac{|F_{\rm M}-F_{\rm A}|}{N_{\rm at}}.
\label{eq:energy-agreement}
\end{equation}
We also report the signed difference per cell $\Delta F=F_{\rm M}-F_{\rm A}$. For convergence trajectories, $F^{(\ell)}$ denotes the free energy of the run at iteration $\ell$.

The iteration count for density-only reinjection gives an empirical reference for the saving achieved by MALA,
\begin{equation}
\eta_N=\frac{N_{\rm A}-N_{\rm M}}{N_{\rm A}-N_{\rm R}},
\label{eq:reinjection-score}
\end{equation}
where $N_{\rm R}$ is the iteration count for reinjection. In all QE cases, $N_{\rm A}>N_{\rm R}$. We report $\eta_N$ in percent.

\section{Results and Discussion}
\label{sec:results}

The more accurate MALA starting densities reduce the SCF iteration count in nine of the 12 matched pairs and tie ATOMIC in the other three. The largest gains occur in the pristine cells, where MALA captures 23.08--37.50\% of the iteration saving observed with density-only reinjection (Table~\ref{tab:iterations}).

\subsection{Initial density accuracy and its spatial distribution}
\label{sec:initial-results}

MALA starts closer to the reference density in all 12 test pairs, both in the relative $L^2$ norm and in the RMS error of the Hartree potential (Table~\ref{tab:global-response}). In pristine B298, the relative $L^2$ error falls from 14.931\% with ATOMIC to 1.023\% with MALA. As vacancies are introduced, the MALA error increases with the number of removed atoms.

MALA lowers the initial RMSE in all three regions defined by distance to the nearest nucleus (Appendix~\ref{app:errors}, Table~\ref{tab:initial-regions}). Around vacancy sites, the trend depends on distance. Within 1.5 \AA\ of a vacancy, MALA gives the larger relative $L^2$ error in all nine boron systems with vacancies. The errors range from 16.46 to 22.01\% with MALA and from 13.60 to 16.47\% with ATOMIC (Table~\ref{tab:vacancy-regions}). MALA again gives the smaller error beyond 1.5 \AA.

\begin{table*}[tb]
\caption{Global errors relative to the converged ATOMIC-initialized QE valence density $n_{\rm ref}$ of each configuration. M and A denote MALA and ATOMIC. The relative density error $\varepsilon_2$ is normalized by $\|n_{\rm ref}\|_2$ and given in percent. Real-space norms retain the zero mode. $\bar{v}_{\mathrm H}$ is the zero-mean Hartree-potential-error RMS in eV, with $\mathbf G=0$ excluded. QE's estimated SCF accuracy and total-energy error are distinct quantities. Initial refers to the native starting density and First to the raw output before mixing, at finite Davidson accuracy. Bold marks the smaller value in each pair and both values at a tie to the displayed precision.}
\label{tab:global-response}
\small
\setlength{\tabcolsep}{3.5pt}
\begin{ruledtabular}
\begin{tabular}{lrrrrrrrr}
& \multicolumn{4}{c}{$\varepsilon_2$ (\%)} & \multicolumn{4}{c}{$\bar{v}_{\mathrm H}$ (eV)} \\
& \multicolumn{2}{c}{Initial} & \multicolumn{2}{c}{First} & \multicolumn{2}{c}{Initial} & \multicolumn{2}{c}{First} \\
Case & M & A & M & A & M & A & M & A \\
\hline
pristine Al298 & \textbf{0.742} & 10.947 & \textbf{0.940} & 2.303 & \textbf{0.068} & 0.443 & 1.634 & \textbf{1.244} \\
pristine Al933 & \textbf{2.021} & 11.504 & 5.630 & \textbf{4.292} & \textbf{0.413} & 0.540 & 10.309 & \textbf{2.503} \\
pristine B298 & \textbf{1.023} & 14.931 & \textbf{1.085} & 7.268 & \textbf{0.261} & 2.025 & \textbf{1.071} & 2.052 \\
B298, 1 vacancy, set 1 & \textbf{2.382} & 14.859 & 10.434 & \textbf{7.307} & \textbf{1.677} & 2.002 & 17.970 & \textbf{2.066} \\
B298, 1 vacancy, set 2 & \textbf{2.428} & 14.857 & 9.492 & \textbf{7.481} & \textbf{1.560} & 2.024 & 16.668 & \textbf{2.393} \\
B298, 1 vacancy, set 3 & \textbf{2.554} & 14.865 & 10.483 & \textbf{7.417} & \textbf{1.679} & 2.021 & 17.888 & \textbf{2.391} \\
B298, 2 vacancies, set 1 & \textbf{3.222} & 14.799 & 9.689 & \textbf{7.528} & \textbf{1.637} & 2.000 & 10.818 & \textbf{2.429} \\
B298, 2 vacancies, set 2 & \textbf{3.173} & 14.806 & 8.240 & \textbf{7.661} & \textbf{1.468} & 2.007 & 7.283 & \textbf{2.901} \\
B298, 2 vacancies, set 3 & \textbf{3.289} & 14.766 & 10.135 & \textbf{7.664} & \textbf{1.724} & 1.996 & 10.108 & \textbf{2.750} \\
B298, 4 vacancies, set 1 & \textbf{4.331} & 14.644 & 10.624 & \textbf{7.770} & \textbf{1.854} & 1.969 & 10.297 & \textbf{2.535} \\
B298, 4 vacancies, set 2 & \textbf{4.230} & 14.645 & 10.600 & \textbf{7.861} & \textbf{1.804} & 1.971 & 11.063 & \textbf{2.763} \\
B298, 4 vacancies, set 3 & \textbf{4.297} & 14.664 & 10.031 & \textbf{7.821} & \textbf{1.760} & 1.979 & 8.945 & \textbf{2.557} \\
\end{tabular}
\end{ruledtabular}
\end{table*}

\subsection{First response and reciprocal-space errors}
\label{sec:response-results}

The smaller initial error of MALA does not always persist through the first SCF step. Before mixing, the global error grows by factors of 1.06--4.38 from the MALA guess and decreases from the ATOMIC guess in every configuration (Table~\ref{tab:global-response}). After this step, the MALA output remains closer to the converged density in Al298 and pristine B298. The ATOMIC output is closer in Al933 and all nine boron systems with vacancies.

The change in the global error reflects a spatially uneven response. In boron systems with vacancies, the first MALA output has larger errors near nuclei and vacancies, while its interstitial RMSE remains smaller than that of ATOMIC (Tables~\ref{tab:first-regions} and \ref{tab:vacancy-regions}). After the first MALA step, the RMS error in the Hartree potential grows by factors of 4.11--24.98 and exceeds the corresponding ATOMIC error in 11 of 12 cases (Table~\ref{tab:global-response}). The spectra and directional maps in Appendix~\ref{app:errors} show error growth at long wavelengths, where the Hartree kernel gives density errors greater weight. The first-step error alone leaves the subsequent convergence path undetermined.

\subsection{Complete SCF convergence and endpoint agreement}
\label{sec:convergence-results}

\begin{figure*}[!t]
\centering
\includegraphics[width=0.95\textwidth]{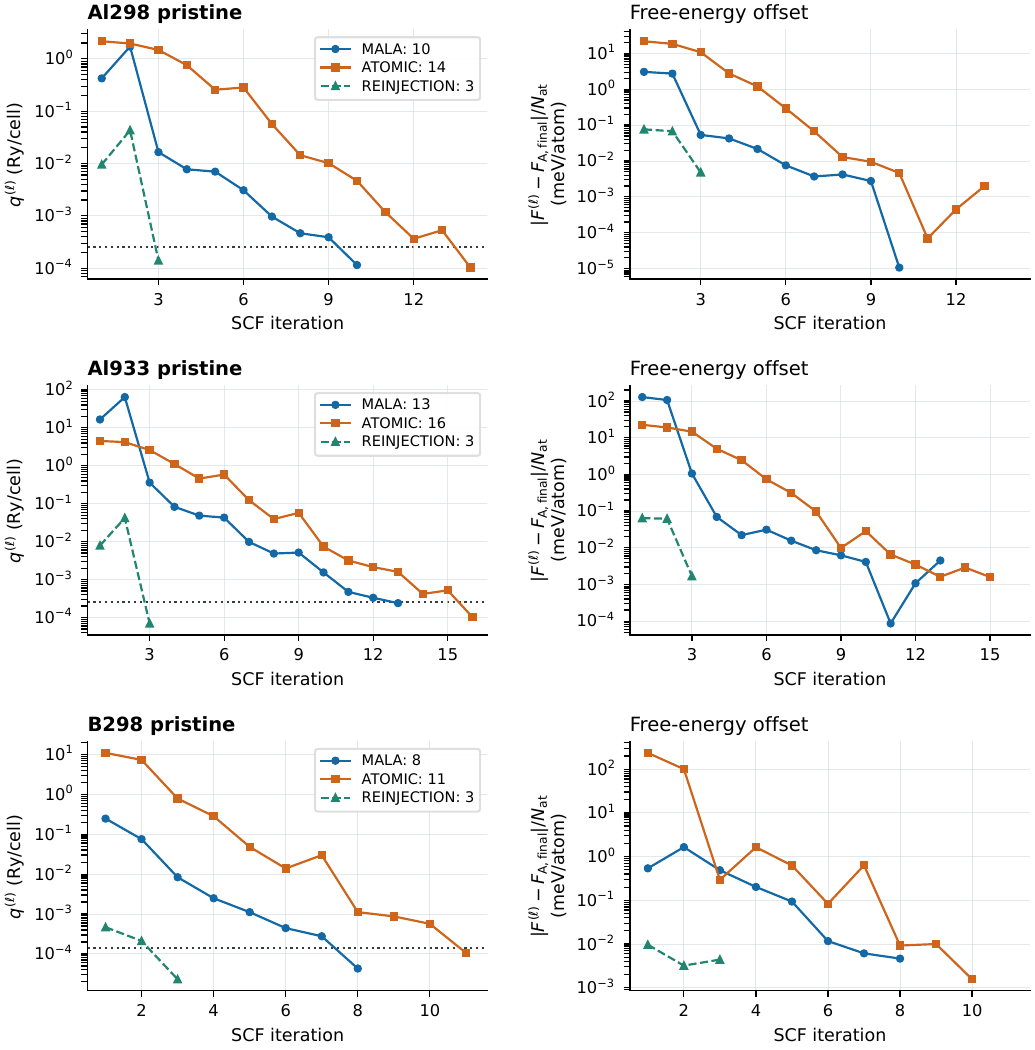}
\caption{QE trajectories for the pristine Al298, Al933, and B298 cells, from top to bottom. MALA (blue), ATOMIC (orange), and density-only reinjection (green dashed) run to convergence. Left panels give QE's estimated SCF accuracy $q^{(\ell)}$ in Ry/cell, a quantity distinct from the real-space density error and the Hartree-potential-error RMS. Dotted lines mark the actual \texttt{conv\_thr}. Right panels give the absolute free-energy offset $|F^{(\ell)}-F_{\rm A}|/N_{\rm at}$ in meV per remaining atom. The axis label $F_{\rm A,final}$ denotes $F_{\rm A}$, the converged ATOMIC free energy for the same configuration. We use $1$ Ry $=13605.693122994$ meV. The logarithmic axis omits exact zeros, including the ATOMIC endpoint, and retains finite nonzero offsets. Vertical scales vary by panel. The count includes any Davidson retry within its outer SCF cycle and excludes the forced-stop first-response replays.}
\label{fig:scf-pristine}
\end{figure*}

All 24 production runs reach the prescribed thresholds. Across the 12 configurations, the density-only reinjection controls converge in two or three iterations and provide the empirical reference for $\eta_N$ (Table~\ref{tab:iterations}). Figure~\ref{fig:scf-pristine} shows the pristine trajectories. In Al933, MALA starts closer to the reference density and has a larger error than ATOMIC after the first unmixed step (Table~\ref{tab:global-response}). The MALA trajectory still reaches the threshold in 13 iterations, compared with 16 for ATOMIC.

In boron systems with vacancies, the savings depend on the placement. The median reductions are two, one, and zero iterations for one, two, and four vacancies, respectively. The divacancies illustrate this variation. Their initial MALA errors span 3.173--3.289\%, and the savings range from zero to three iterations. Comparisons with pristine B also reflect its looser stopping threshold. Appendix~\ref{app:parameters} gives the individual vacancy trajectories in Figs.~\ref{fig:scf-one-vacancy}--\ref{fig:scf-four-vacancies} and summarizes the iteration savings by vacancy count in Table~\ref{tab:vacancy-control-groups}.

Despite the different paths, the absolute free-energy difference $\Delta e$ is at most $0.00465$ meV/atom, and the final relative density error $\varepsilon_{2,\rm final}$ is at most 0.0365\% at the prescribed tolerances. Appendix~\ref{app:parameters} gives the production endpoint comparisons in Table~\ref{tab:endpoints}. Tables~\ref{tab:reinjection-endpoints} and \ref{tab:reinjection-density-endpoints} give the energy and density checks for the reinjection controls, respectively.

\begin{table*}[tb]
\centering
\caption{QE iteration counts for all 12 configurations. M, A, and R denote MALA, ATOMIC, and reinjection of the converged native density of the same configuration. The saving is $\Delta N=N_{\rm A}-N_{\rm M}$. We report $\eta_N$, the fraction of the iteration saving observed with reinjection, in percent [Eq.~\eqref{eq:reinjection-score}]. Bold compares M and A, with both bold at a tie. Each entry counts outer iterations to convergence for one trajectory, including the first and excluding forced one-step diagnostics. The thresholds are $2.56\times10^{-4}$ Ry/cell for Al, $1.44\times10^{-4}$ for pristine B, and $10^{-6}$ for boron with vacancies. They match within each case and contribute to differences between pristine boron and boron with vacancies.}
\label{tab:iterations}
\small
\setlength{\tabcolsep}{4pt}
\begin{tabular}{lrrrrr}
\hline
Case & $N_{\rm M}$ & $N_{\rm A}$ & $\Delta N$ & $N_{\rm R}$ & $\eta_N$ (\%) \\
\hline
pristine Al298 & \textbf{10} & 14 & 4 & 3 & 36.36 \\
pristine Al933 & \textbf{13} & 16 & 3 & 3 & 23.08 \\
pristine B298 & \textbf{8} & 11 & 3 & 3 & 37.50 \\

B298, 1 vacancy, set 1 & \textbf{15} & 16 & 1 & 2 & 7.14 \\
B298, 1 vacancy, set 2 & \textbf{15} & 17 & 2 & 3 & 14.29 \\
B298, 1 vacancy, set 3 & \textbf{16} & 18 & 2 & 3 & 13.33 \\

B298, 2 vacancies, set 1 & \textbf{18} & 21 & 3 & 3 & 16.67 \\
B298, 2 vacancies, set 2 & \textbf{21} & 22 & 1 & 3 & 5.26 \\
B298, 2 vacancies, set 3 & \textbf{21} & \textbf{21} & 0 & 3 & 0.00 \\

B298, 4 vacancies, set 1 & \textbf{17} & 18 & 1 & 2 & 6.25 \\
B298, 4 vacancies, set 2 & \textbf{18} & \textbf{18} & 0 & 2 & 0.00 \\
B298, 4 vacancies, set 3 & \textbf{17} & \textbf{17} & 0 & 2 & 0.00 \\
\hline
\end{tabular}
\end{table*}

\section{Conclusions}
\label{sec:conclusions}

We supply MALA densities directly to the plane-wave code QE for a small set of aluminum and boron configurations. MALA reduces the SCF iteration count in nine of the 12 matched pairs and ties ATOMIC in the other three. The final valence densities and free energies agree closely at the prescribed thresholds. The largest savings occur in pristine cells. For one, two, and four vacancies, the mean reinjection-referenced fractions $\eta_N$ are 11.59\%, 7.31\%, and 2.08\%, respectively. 

MALA lowers the global starting-density error even though it gives larger errors close to vacancies. The first unmixed step increases the global MALA error, and Fourier analysis locates part of this growth at long wavelengths, where density errors contribute strongly to the Hartree potential. Finite diagonalization accuracy affects the first response, and mixing shapes the subsequent trajectory.

The free-energy trajectories illustrate how a more accurate starting density can bring the calculation close to the converged energy in fewer iterations. These results motivate the development of density predictors that transfer across materials and structures, with the longer-term goal of a universal initializer for routine Kohn--Sham DFT calculations.

\section*{Acknowledgments}

This work was supported by the Center for Advanced Systems Understanding (CASUS), which is financed by Germany’s Federal Ministry of Research, Technology and Space (BMFTR) and by the Saxon State government out of the State budget approved by the Saxon State Parliament. Computational resources were provided by the \texttt{rosi} clusters of HZDR.

\appendix

\section{Transfer of the predicted density to QE}
\label{app:qe-transfer}

The two initializers differ in how they supply the density for the first QE potential. We select the ATOMIC guess with \texttt{startingpot} set to \texttt{'atomic'} and read the MALA density from the save directory with \texttt{'file'}~\cite{QEInputPW}. This QE build stores the Fourier coefficients in \texttt{charge-density.hdf5} and the structure and grid metadata in Extensible Markup Language (XML) format. In both cases, \texttt{restart\_mode='from\_scratch'} and \texttt{startingwfc='atomic+random'} instruct QE to construct fresh wave functions.

For reinjection, we copy the converged ATOMIC \texttt{charge-density.hdf5} unchanged, bypassing conversion and interpolation, and switch \texttt{startingpot} to \texttt{'file'}. We transfer the density alone and leave the wave functions and XML electronic restart data behind. The run inputs and the hashes of the source densities document the transfer.

To represent the MALA density on the native QE grid, we take the FFT grid, Miller indices, and reciprocal-vector order from a template for the same configuration. We replace its electronic coefficients with those of the MALA density. The resulting file defines the initial MALA field, with the template supplying the representation and with no fit to a converged density. The analysis records verify the geometry, electron count, grid, origin, and ordering. Field paths and hashes identify the density at each stage.

The real-space densities are tabulated in $e\,\text{\AA}^{-3}$, whereas the native coefficients use $e\,a_0^{-3}$, where $e$ is the elementary charge and $a_0$ is the Bohr radius. With $a_0=0.529177210903$ \AA, we multiply the density by $a_0^3$ to obtain native units and divide by the same factor to recover the tabulated units. We normalize the discrete Fourier transform by $1/N_g$, where $N_g$ is the number of real-space grid points. The inverse is a Fourier sum. At $\Gamma$, QE stores one vector from each nonzero conjugate pair, so we recover its partner by complex conjugation when reconstructing the real-space density. Multiplying the zero coefficient by the cell volume in matching units gives the electron count.

The converter and the 12 import audits check the source cell, geometry, complete array coverage, units, electron count, and template metadata. The source and target grids have equal dimensions, so all 12 imports use the original real-space grid without resampling. Projection onto the native reciprocal sphere still changes the density by the amounts in Table~\ref{tab:native-projection}. We read the coefficients back from the file and compare them with the intended values. The small support archive records hashes linking the source arrays to the imported fields and omits the large arrays themselves.

The converter leaves the density unclipped and unnormalized and performs no fit to the converged reference. The model restricts negative LDOS at the earlier prediction stage (Sec.~\ref{sec:mala}). The projection difference compares the raw MALA density with its native representation and uses neither a converged reference nor an SCF count. Because we evaluate the errors of the imported field, those errors include the effect of the projection. QE applies its usual normalization to the ATOMIC guess.

The production runs did not save the first unmixed densities. To recover them in separate replays, we retain the QE 7.5 executable and the initial diagonalization tolerance \texttt{diago\_thr\_init}, lower \texttt{electron\_maxstep} from 100 to 1, set \texttt{scf\_must\_converge} to false, and change \texttt{disk\_io} from \texttt{low} to \texttt{nowf}. The forced stop saves the unmixed density through the existing QE output path and prints a convergence message after one iteration. Sixteen replays retain Davidson warnings for one to three unconverged eigenvalues, and the occupations of the states that trigger the warnings remain undetermined (Appendix~\ref{app:parameters}). Because the forced endpoint evaluates the energy differently, we use these runs only to diagnose the first density response.

\begin{table*}[tb]
\caption{Native QE density transfer for the 12 configurations. The relative projection difference $\|n_{\rm native}-n_{\rm source}\|_2/\|n_{\rm source}\|_2$ compares the MALA density before export, $n_{\rm source}$, with its reconstruction on QE's reciprocal sphere, $n_{\rm native}$, in the same units. The difference is given in percent. The Fourier projection can truncate components even when the real-space grids have the same dimensions. The comparison quantifies the transfer independently of initializer accuracy and uses no converged ATOMIC reference. $N_e$ is the imported valence-electron count rounded to an integer, with agreement within floating-point precision.}
\label{tab:native-projection}
\small
\setlength{\tabcolsep}{3.5pt}
\begin{ruledtabular}
\begin{tabular}{lrr}
Configuration & Projection difference (\%) & $N_e$ \\
\hline
pristine Al298 & 0.112249 & 768 \\
pristine Al933 & 0.167178 & 768 \\
pristine B298 & 0.143789 & 432 \\
B298, 1 vacancy, set 1 & 0.143977 & 429 \\
B298, 1 vacancy, set 2 & 0.143962 & 429 \\
B298, 1 vacancy, set 3 & 0.143790 & 429 \\
B298, 2 vacancies, set 1 & 0.144654 & 426 \\
B298, 2 vacancies, set 2 & 0.144186 & 426 \\
B298, 2 vacancies, set 3 & 0.144340 & 426 \\
B298, 4 vacancies, set 1 & 0.145804 & 420 \\
B298, 4 vacancies, set 2 & 0.145505 & 420 \\
B298, 4 vacancies, set 3 & 0.145497 & 420 \\
\end{tabular}
\end{ruledtabular}
\end{table*}

\section{Computational parameters and convergence checks}
\label{app:parameters}

Table~\ref{tab:qe-parameters} lists the settings for the production pairs. We use PBEsol with \texttt{Al.sr-pbesol.upf} for aluminum and PBE with \texttt{B\_ncrs\_v041\_pbe.upf} for boron. The production logs identify both NC potentials as generated with ONCVPSP, with three valence electrons and a core correction. QE includes the correction in the Hamiltonian, and we evaluate the density errors using only the valence density.

\begin{table*}[tb]
\caption{Numerical settings for the paired QE production runs, which differ only in the density initializer. We use neutral cells with \texttt{nspin=1} and Fermi--Dirac occupations. All inputs set \texttt{nosym=.true.}, with \texttt{noinv} false for Al and true for B. Every pair uses \texttt{diagonalization='david'}, starts with \texttt{startingwfc='atomic+random'} and \texttt{restart\_mode='from\_scratch'}, and uses \texttt{disk\_io='low'}. For plain modified-Broyden mixing, we set \texttt{mixing\_beta=0.1} and \texttt{mixing\_ndim=8}. The wave-function and charge-density cutoffs $E_{\rm wfc}$ and $E_\rho$ are in Ry, and \texttt{conv\_thr} sets the estimated SCF energy-error threshold per cell. The $8^3$ Al and $4^3$ B meshes refer to the training calculations. The execution layout is given in Message Passing Interface (MPI) ranks and threads per rank.}
\label{tab:qe-parameters}
\footnotesize
\setlength{\tabcolsep}{3.5pt}
\begin{ruledtabular}
\begin{tabular}{lrrrr}
Parameter & Al298 & Al933 & pristine B298 & B298 with vacancies \\
\hline
QE version & 7.5 & 7.5 & 7.5 & 7.5 \\
Exchange-correlation & PBEsol & PBEsol & PBE & PBE \\
$E_{\rm wfc}/E_\rho$ (Ry) & 100/400 & 100/400 & 80/320 & 80/320 \\
FFT grid & $200^3$ & $200^3$ & $108\times108\times135$ & $108\times108\times135$ \\
$\mathbf k$ sampling & $\Gamma$ & $\Gamma$ & $\Gamma$ & $\Gamma$ \\
Number of bands & 576 & 576 & 280 & 280 \\
Fermi--Dirac \texttt{degauss} (Ry) & 0.0019 & 0.0059 & 0.0018874 & 0.0018874 \\
\texttt{conv\_thr} (Ry/cell) & $2.56\times10^{-4}$ & $2.56\times10^{-4}$ & $1.44\times10^{-4}$ & $10^{-6}$ \\
\texttt{diago\_thr\_init} (Ry) & $10^{-5}$ & $10^{-5}$ & $10^{-5}$ & $10^{-5}$ \\
\texttt{electron\_maxstep} & 100 & 100 & 100 & 100 \\
MPI ranks $\times$ threads/rank & $8\times1$ & $8\times1$ & $4\times1$ & $4\times1$ \\
Allocated memory (GiB) & 40 & 40 & 4 & 4 \\
\end{tabular}
\end{ruledtabular}
\end{table*}

Both aluminum configurations share a cubic cell with side $16.19800005$ \AA\ after rounding. The real-space lattice vectors $\mathbf a_1$, $\mathbf a_2$, and $\mathbf a_3$ of the boron cell are, in \AA,
\begin{equation}
\begin{split}
\mathbf a_1&=(9.79678558,\,0,\,0),\\
\mathbf a_2&=(-4.89839279,\,8.484265188,\,0),\\
\mathbf a_3&=(0,\,0,\,12.52416131).
\end{split}
\end{equation}
We keep these vectors and the coordinates of the remaining B298 atoms in every boron system with vacancies. The configuration records identify the geometry and sites (Table~\ref{tab:benchmark-systems}). For the numerical settings, including \texttt{K\_POINTS gamma}, we use the production inputs and logs as the record, independently of the dense-mesh planning metadata.

We run the production pairs on \texttt{cpu-genoa-int} with physical-core binding and single-thread numerical libraries. Each run uses eight MPI ranks for Al or four for B, with one OpenMP thread per rank. We use the same layout for both runs in each pair.

\subsection{Model and execution provenance}

The public aluminum archive identifies MALA~1.2.1 as the version used for retraining~\cite{fiedler2024rodare2725}. The later Al298 prediction used MALA~1.4.0. The original boron training environment and the full package manifests for some earlier predictions remain unknown. The current source and application programming interface documentation describe the workflow, and contemporaneous records establish the historical versions where available.

The layer widths are $91\to800\to800\to800\to250\to250$ for Al298, $91\to4000\to4000\to4000\to250\to250$ for Al933, and $91\to800\to800\to800\to250\to241$ for boron. The archive identity, tensor dimensions, and source-density audit link the Al933 architecture to the starting density for snapshot 9. The public solid-933 model also has 4000-wide layers. The physical role of that width remains untested because we use fixed architectures. The audit verifies the network identity independently of the unresolved liquid-only training composition.

The Al298 prediction used MALA~1.4.0, PyTorch~2.14.0, NumPy~2.5.3, SciPy~1.18.1, ASE~3.29.0, and mpi4py~4.1.2. The network arithmetic used float32, and the assembled LDOS used float64. The boron vacancy worker likewise verifies float32 tensors on the central processing unit (CPU) and float64 LDOS. The compatibility record distinguishes this precision from the float32 LDOS used in the earlier pristine calculations. In the current MALA source, the algorithm brackets the electron-number equation over the LDOS energy interval, solves it with \texttt{toms748}, and integrates analytically over energy. The contemporaneous software manifests for the older Al933 and boron predictions remain incomplete.

The QE runtime records identify AMD EPYC~9654 resources, QE~7.5, GCC~14.3.0, and OpenMPI~5.0.8. The eight ranks for Al and four for B each use one thread, and the Basic Linear Algebra Subprograms library is also limited to one thread. The Al298 density-generation worker uses eight MPI ranks for the descriptors, eight Torch threads on the root rank, and CPU inference. For the divacancy worker, the records show eight CPU ranks on AMD EPYC~9654, one thread per rank, and paths for GCC~14.2.0, OpenMPI~4.1.8, and Python~3.12.4. These records establish the devices and threading for the identified jobs.

The reproducibility records identify the QE executable and the converter and bridge sources by their SHA-256 hashes. The QE library paths identify OpenBLAS~0.3.30, FlexiBLAS~3.4.5, FFTW~3.3.10, ELPA~2025.06.002, and HDF5~1.14.6. The archived Al and B UPF files verify the pseudopotential identities. Together, these records identify the implementation and execution layout.

\begin{table*}[tb]
\caption{Labels and run identifiers for the 12 configurations. Set numbers distinguish placements at each vacancy count, with no nesting implied. The zero-based indices of the removed sites in the parent cell are given in Table~\ref{tab:benchmark-systems}.}
\label{tab:case-identifiers}
\small
\setlength{\tabcolsep}{3.5pt}
\begin{ruledtabular}
\begin{tabular}{ll}
Manuscript label & Internal identifier \\
\hline
pristine Al298 & \texttt{Al298} \\
pristine Al933 & \texttt{Al933} \\
pristine B298 & \texttt{B298} \\
B298, 1 vacancy, set 1 & \texttt{B298\_vac02} \\
B298, 1 vacancy, set 2 & \texttt{B298\_vac03} \\
B298, 1 vacancy, set 3 & \texttt{B298\_vac06} \\
B298, 2 vacancies, set 1 & \texttt{B298\_dv2\_c1} \\
B298, 2 vacancies, set 2 & \texttt{B298\_dv2\_c2} \\
B298, 2 vacancies, set 3 & \texttt{B298\_dv2\_c3} \\
B298, 4 vacancies, set 1 & \texttt{B298\_mvac04} \\
B298, 4 vacancies, set 2 & \texttt{B298\_v4\_c1} \\
B298, 4 vacancies, set 3 & \texttt{B298\_v4\_c2} \\
\end{tabular}
\end{ruledtabular}
\end{table*}

\subsection{QE convergence and endpoint checks}

Table~\ref{tab:endpoints} compares the production runs for all 12 configurations through the signed free-energy differences and the final density errors. The largest absolute free-energy differences are $0.004640$ meV/atom for pristine B and $0.004508$ meV/atom for Al933. The other cases differ by at most $2.00\times10^{-5}$ meV/atom. The relative $L^2$ differences between the final densities range from 0.0338 to 0.0365\% for the pristine cells and from 0.00105 to 0.00257\% for boron systems with vacancies. The comparison between these ranges also reflects the looser thresholds for the pristine cells.

Reinjection converges in three iterations for each pristine cell. With those counts as the reference, MALA saves four, three, and three of the iterations between ATOMIC and reinjection for Al298, Al933, and B298, giving $\eta_N=36.36\%$, 23.08\%, and 37.50\%, respectively. The reinjection counts for the single-vacancy cells are 2, 3, and 3. The divacancies take three iterations each, and the four-vacancy cells take two. The corresponding ranges of $\eta_N$ are 7.14--14.29\%, 0--16.67\%, and 0--6.25\%. Table~\ref{tab:vacancy-control-groups} summarizes the selected placements in two ways. The mean averages the individual fractions, and the pooled fraction divides the sum of the savings by the sum of the ATOMIC counts above the controls.

In each vacancy control, QE retries Davidson diagonalization at a tighter threshold within the first outer iteration. The iteration count measures the work of reaching convergence again with the same settings and fresh orbitals. The density-import error and the minimum attainable count remain undetermined. The final energies and densities agree closely with the ATOMIC endpoints. For the nine vacancy controls, the largest free-energy difference is $3.4\times10^{-7}$ Ry/cell and the largest relative $L^2$ density difference is $3.58\times10^{-5}$ ($0.00358\%$). Tables~\ref{tab:reinjection-endpoints} and \ref{tab:reinjection-density-endpoints} give the endpoint checks.

For pristine Al298, Al933, and B298, the occupations of the highest band are approximately $8.26\times10^{-66}$, $3.61\times10^{-19}$, and $3.65\times10^{-81}$ with ATOMIC, and $9.46\times10^{-66}$, $3.65\times10^{-19}$, and $3.61\times10^{-81}$ with reinjection, respectively. These values establish that the top bands are empty. The occupations of the states that trigger a first Davidson warning remain undetermined. The endpoint comparisons in Tables~\ref{tab:endpoints} and \ref{tab:reinjection-endpoints} apply at the stated finite thresholds and leave exact state identity untested.

\begin{table*}[!tbp]

\caption{Final QE free energies and valence-density errors for the converged production pairs. M and A identify the starting densities. QE logs the free energy $F=E-T_e S$ as \texttt{! total energy} in Ry/cell. The signed difference is $\Delta F=F_{\rm M}-F_{\rm A}$, with $\Delta F/N_{\rm at}$ expressed per remaining atom using $1$ Ry $=13605.693122994$ meV. The density error $\varepsilon_{2,\rm final}=\|n_{\rm M,final}-n_{\rm A,final}\|_2/\|n_{\rm A,final}\|_2$ is given in percent for densities on the common native grid. The comparisons retain the finite SCF tolerance and the printed energy precision.}
\label{tab:endpoints}
\footnotesize
\setlength{\tabcolsep}{3.5pt}
\begin{ruledtabular}
\begin{tabular}{lrrrrr}
Case & $F_{\rm M}$ (Ry/cell) & $F_{\rm A}$ (Ry/cell) & $\Delta F$ (Ry/cell) & $\Delta F/N_{\rm at}$ (meV) & $\varepsilon_{2,\rm final}$ (\%) \\
\hline
pristine Al298 & $-1187.06623963$ & $-1187.06623943$ & $-2\times10^{-7}$ & $-1.063\times10^{-5}$ & 0.03447 \\
pristine Al933 & $-1183.71049462$ & $-1183.71057945$ & $+8.483\times10^{-5}$ & $+4.508\times10^{-3}$ & 0.03652 \\
pristine B298 & $-846.41157080$ & $-846.41152169$ & $-4.911\times10^{-5}$ & $-4.64\times10^{-3}$ & 0.03379 \\
B298, 1 vacancy, set 1 & $-840.30056365$ & $-840.30056383$ & $+1.8\times10^{-7}$ & $+1.713\times10^{-5}$ & 0.00237 \\
B298, 1 vacancy, set 2 & $-840.19648703$ & $-840.19648682$ & $-2.1\times10^{-7}$ & $-1.998\times10^{-5}$ & 0.00257 \\
B298, 1 vacancy, set 3 & $-840.18464867$ & $-840.18464848$ & $-1.9\times10^{-7}$ & $-1.808\times10^{-5}$ & 0.00251 \\
B298, 2 vacancies, set 1 & $-834.10078864$ & $-834.10078868$ & $+4\times10^{-8}$ & $+3.833\times10^{-6}$ & 0.00199 \\
B298, 2 vacancies, set 2 & $-834.10742601$ & $-834.10742591$ & $-1\times10^{-7}$ & $-9.581\times10^{-6}$ & 0.00116 \\
B298, 2 vacancies, set 3 & $-834.08352888$ & $-834.08352890$ & $+2\times10^{-8}$ & $+1.916\times10^{-6}$ & 0.00105 \\
B298, 4 vacancies, set 1 & $-821.83157320$ & $-821.83157316$ & $-4\times10^{-8}$ & $-3.887\times10^{-6}$ & 0.00204 \\
B298, 4 vacancies, set 2 & $-821.81557525$ & $-821.81557522$ & $-3\times10^{-8}$ & $-2.916\times10^{-6}$ & 0.00214 \\
B298, 4 vacancies, set 3 & $-821.81971442$ & $-821.81971437$ & $-5\times10^{-8}$ & $-4.859\times10^{-6}$ & 0.00187 \\
\end{tabular}
\end{ruledtabular}

\vspace{0pt}

\centering
\caption{Mean iteration counts and fractions for the three placements at each boron vacancy count. We give the counts to three decimal places and the percentages to two. The mean of $\eta_N$ averages the three case-specific ratios. The pooled fraction is $\sum_i(N_{{\rm A},i}-N_{{\rm M},i})/\sum_i(N_{{\rm A},i}-N_{{\rm R},i})$, with $i$ running over the three placements at that vacancy count. Both quantities summarize the selected geometries and leave the population mean, statistical significance, and any universal vacancy-concentration boundary undetermined.}
\label{tab:vacancy-control-groups}
\small
\setlength{\tabcolsep}{5pt}
\begin{tabular}{rrrrrrr}
\hline
Vacancies & Placements & Mean $N_{\rm M}$ & Mean $N_{\rm A}$ & Mean $N_{\rm R}$ & Mean $\eta_N$ (\%) & Pooled $\eta_N$ (\%) \\
\hline
1 & 3 & 15.333 & 17.000 & 2.667 & 11.59 & 11.63 \\
2 & 3 & 20.000 & 21.333 & 3.000 & 7.31 & 7.27 \\
4 & 3 & 17.333 & 17.667 & 2.000 & 2.08 & 2.13 \\
\hline
\end{tabular}

\vspace{0pt}

\centering
\caption{Endpoints of the 12 QE density-only reinjection controls. The table gives the converged QE free energy $F_{\rm R}$, with $F=E-T_e S$, the final value $q_{\rm R}$ of the estimated SCF accuracy $q^{(\ell)}$, and the signed difference $\Delta F_{\rm R}=F_{\rm R}-F_{\rm A}$ from the original ATOMIC endpoint of the same configuration, all in Ry/cell. $N_{\rm R}$ counts outer iterations to convergence at the thresholds in Table~\ref{tab:qe-parameters}. The differences reflect the finite stopping tolerances and printed precision. They establish neither state identity nor an energy-error bound, and a more negative endpoint provides no ranking of the initializers.}
\label{tab:reinjection-endpoints}
\small
\setlength{\tabcolsep}{5pt}
\begin{tabular}{lrrrr}
\hline
Case & $N_{\rm R}$ & $F_{\rm R}$ (Ry/cell) & $q_{\rm R}$ (Ry/cell) & $\Delta F_{\rm R}$ (Ry/cell) \\
\hline
pristine Al298 & 3 & $-1187.06614919$ & $1.4066\times10^{-4}$ & $+9.024\times10^{-5}$ \\
pristine Al933 & 3 & $-1183.71054761$ & $6.974\times10^{-5}$ & $+3.184\times10^{-5}$ \\
pristine B298 & 3 & $-846.41156830$ & $2.319\times10^{-5}$ & $-4.661\times10^{-5}$ \\

B298, 1 vacancy, set 1 & 2 & $-840.30056349$ & $7.5\times10^{-7}$ & $+3.4\times10^{-7}$ \\
B298, 1 vacancy, set 2 & 3 & $-840.19648713$ & $1.8\times10^{-7}$ & $-3.1\times10^{-7}$ \\
B298, 1 vacancy, set 3 & 3 & $-840.18464873$ & $1.2\times10^{-7}$ & $-2.5\times10^{-7}$ \\

B298, 2 vacancies, set 1 & 3 & $-834.10078867$ & $1.9\times10^{-7}$ & $+1\times10^{-8}$ \\
B298, 2 vacancies, set 2 & 3 & $-834.10742588$ & $2\times10^{-7}$ & $+3\times10^{-8}$ \\
B298, 2 vacancies, set 3 & 3 & $-834.08352877$ & $1\times10^{-7}$ & $+1.3\times10^{-7}$ \\

B298, 4 vacancies, set 1 & 2 & $-821.83157294$ & $5.9\times10^{-7}$ & $+2.2\times10^{-7}$ \\
B298, 4 vacancies, set 2 & 2 & $-821.81557495$ & $6.6\times10^{-7}$ & $+2.7\times10^{-7}$ \\
B298, 4 vacancies, set 3 & 2 & $-821.81971429$ & $3.7\times10^{-7}$ & $+8\times10^{-8}$ \\
\hline
\end{tabular}

\vspace{0pt}

\centering
\caption{Final valence-density differences for the nine vacancy reinjection controls at finite tolerance. The dimensionless error is $\|n_{\rm R,final}-n_{\rm A,final}\|_2/\|n_{\rm A,final}\|_2$ on the common native QE grid. Here $n_{\rm R,final}$ is the converged density from the reinjection run, and $n_{\rm A,final}$ is the converged ATOMIC density used to start it. The audit evaluates the difference by Gamma-weighted Parseval sums with the same Miller ordering and no renormalization. Sets distinguish placements within each vacancy count.}
\label{tab:reinjection-density-endpoints}
\small
\begin{tabular}{rrrr}
\hline
Vacancies & Set 1 & Set 2 & Set 3 \\
\hline
1 & $3.47457\times10^{-5}$ & $3.17027\times10^{-5}$ & $2.82930\times10^{-5}$ \\
2 & $2.24507\times10^{-5}$ & $2.11420\times10^{-5}$ & $1.21534\times10^{-5}$ \\
4 & $3.57782\times10^{-5}$ & $3.54259\times10^{-5}$ & $2.95278\times10^{-5}$ \\
\hline
\end{tabular}

\end{table*}

\begin{figure*}[!t]
\centering
\includegraphics[width=0.95\textwidth]{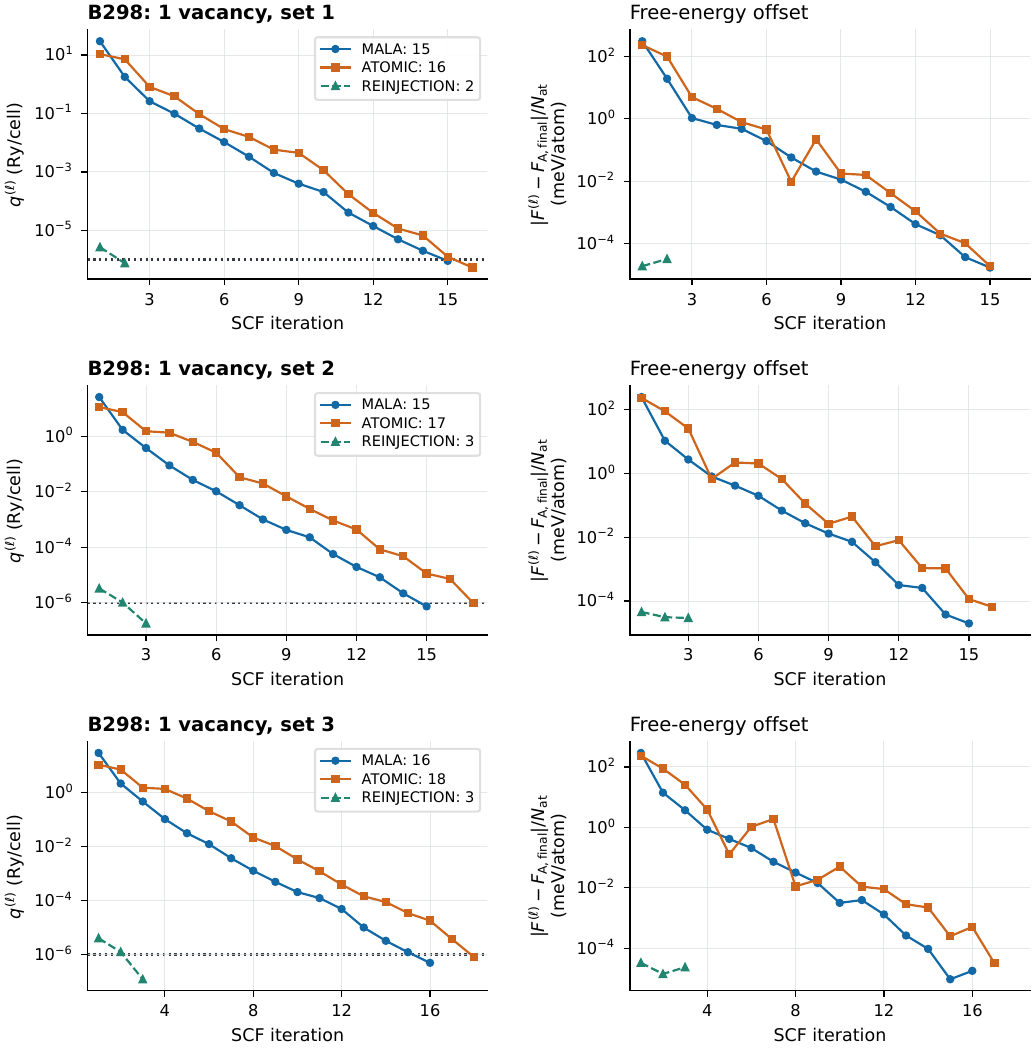}
\caption{QE trajectories for the three single-vacancy B298 cells, sets 1--3 from top to bottom. The curves show MALA (blue), ATOMIC (orange), and density-only reinjection (green dashed) through convergence. In the left panels, $q^{(\ell)}$ is QE's estimated SCF accuracy in Ry/cell, and the dotted lines show the actual \texttt{conv\_thr}. The real-space density error and the Hartree-potential-error RMS are distinct quantities. The right panels give the free-energy offset $|F^{(\ell)}-F_{\rm A}|/N_{\rm at}$ in meV per remaining atom. The axis label $F_{\rm A,final}$ denotes $F_{\rm A}$, the converged ATOMIC free energy for the same configuration. We use $1$ Ry $=13605.693122994$ meV. The absolute value omits the sign. The logarithmic axis displays finite nonzero offsets and omits exact zeros, including the ATOMIC endpoint. The vertical scale varies by panel. Each Davidson retry counts within its outer SCF cycle, and the trajectories exclude the forced-stop first-response replays.}
\label{fig:scf-one-vacancy}
\end{figure*}

\begin{figure*}[!t]
\centering
\includegraphics[width=0.95\textwidth]{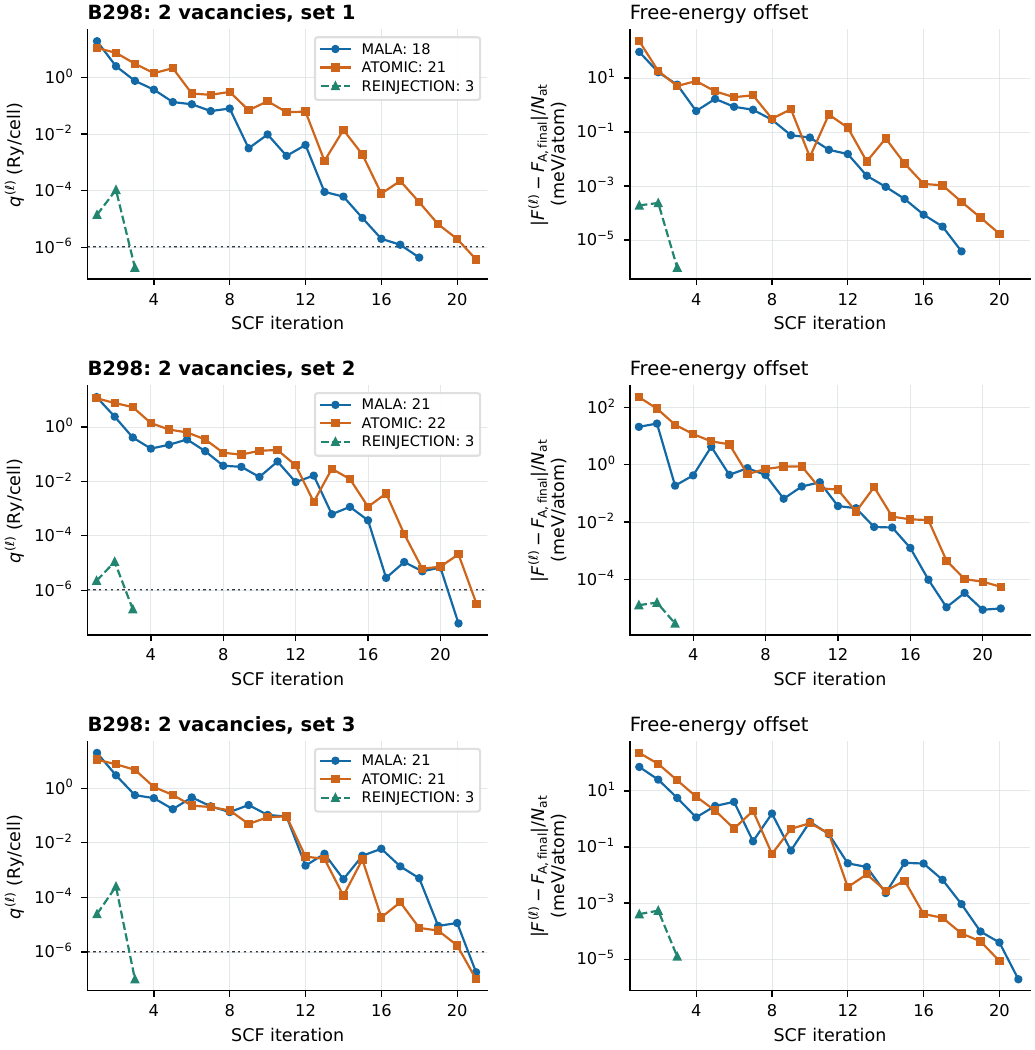}
\caption{QE trajectories for the three divacancy B298 cells, sets 1--3 from top to bottom. The curves show MALA (blue), ATOMIC (orange), and density-only reinjection (green dashed) through convergence. In the left panels, $q^{(\ell)}$ is QE's estimated SCF accuracy in Ry/cell, and the dotted lines show the actual \texttt{conv\_thr}. The real-space density error and the Hartree-potential-error RMS are distinct quantities. The right panels give the free-energy offset $|F^{(\ell)}-F_{\rm A}|/N_{\rm at}$ in meV per remaining atom. The axis label $F_{\rm A,final}$ denotes $F_{\rm A}$, the converged ATOMIC free energy for the same configuration. We use $1$ Ry $=13605.693122994$ meV. The absolute value omits the sign. The logarithmic axis displays finite nonzero offsets and omits exact zeros, including the ATOMIC endpoint. The vertical scale varies by panel. Each Davidson retry counts within its outer SCF cycle, and the trajectories exclude the forced-stop first-response replays.}
\label{fig:scf-two-vacancies}
\end{figure*}

\begin{figure*}[!t]
\centering
\includegraphics[width=0.95\textwidth]{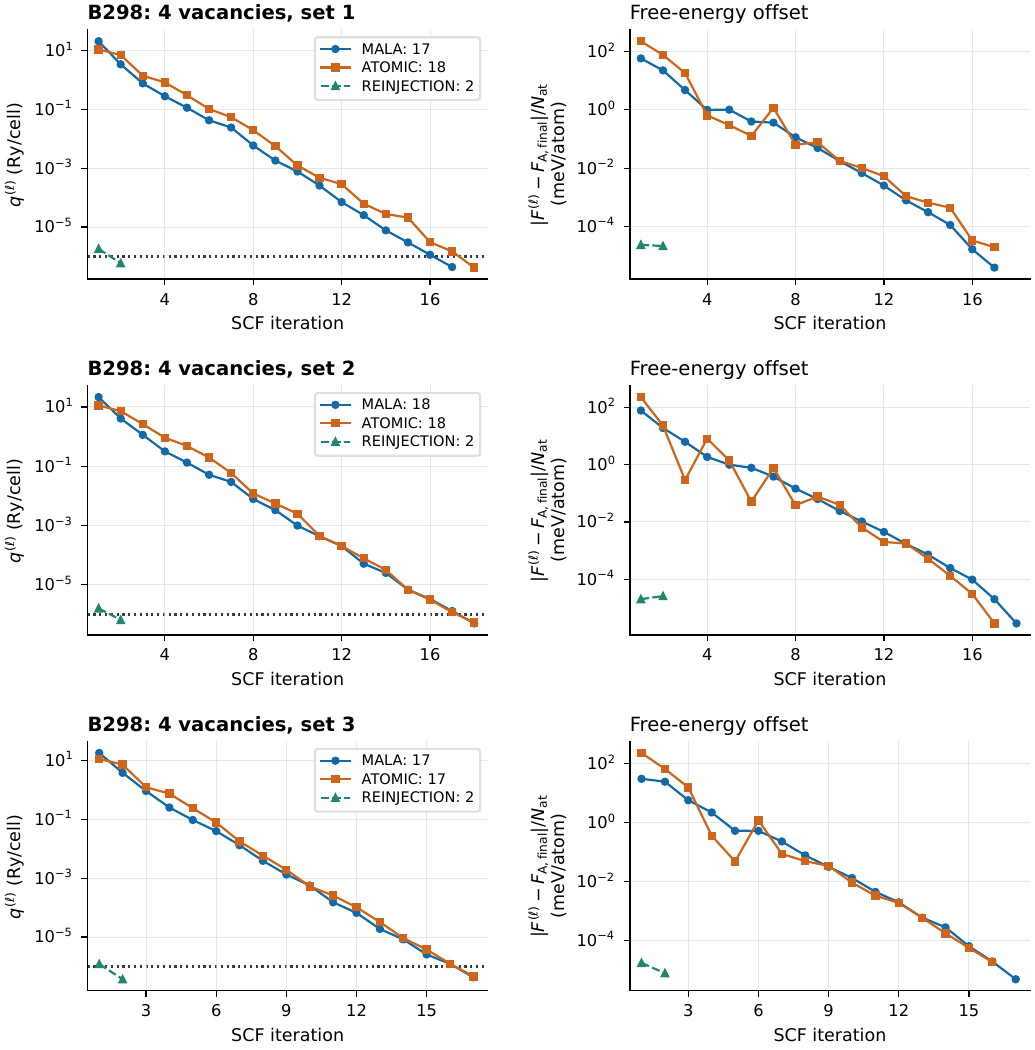}
\caption{QE trajectories for the three four-vacancy B298 cells, sets 1--3 from top to bottom. The curves show MALA (blue), ATOMIC (orange), and density-only reinjection (green dashed) through convergence. In the left panels, $q^{(\ell)}$ is QE's estimated SCF accuracy in Ry/cell, and the dotted lines show the actual \texttt{conv\_thr}. The real-space density error and the Hartree-potential-error RMS are distinct quantities. The right panels give the free-energy offset $|F^{(\ell)}-F_{\rm A}|/N_{\rm at}$ in meV per remaining atom. The axis label $F_{\rm A,final}$ denotes $F_{\rm A}$, the converged ATOMIC free energy for the same configuration. We use $1$ Ry $=13605.693122994$ meV. The absolute value omits the sign. The logarithmic axis displays finite nonzero offsets and omits exact zeros, including the ATOMIC endpoint. The vertical scale varies by panel. Each Davidson retry counts within its outer SCF cycle, and the trajectories exclude the forced-stop first-response replays.}
\label{fig:scf-four-vacancies}
\end{figure*}

Figures~\ref{fig:scf-one-vacancy}--\ref{fig:scf-four-vacancies} show how the convergence trajectories vary among the nine boron systems with vacancies. In set order, MALA saves 1, 2, and 2 iterations for the single-vacancy cells, 3, 1, and 0 for the divacancies, and 1, 0, and 0 for the four-vacancy cells.

For each configuration, the trajectories show $|F^{(\ell)}-F_{\rm A}|/N_{\rm at}$ in meV per remaining atom, using the ATOMIC free energy $F_{\rm A}$ at the production threshold as the reference. The logarithmic axis omits the zero ATOMIC endpoint. The magnitude discards the sign, so Table~\ref{tab:endpoints} gives the signed differences separately. The plotted offset is distinct from a monotonic variational error and leaves both the tighter-reference energy and the identity of the electronic states undetermined.

\clearpage
\section{Error measures and diagnostics}
\label{app:errors}

\subsection{Real-space density errors}

For each configuration, all density fields share the cell, coordinates, native reciprocal vectors, zero origin, and uniform FFT grid with $N_g$ points, each representing a volume $\Delta V$. All 72 saved fields are finite and nonnegative, and their absolute charge errors are below $10^{-10}\,e$. We compare these fields directly, without renormalization or a fitted spatial shift. At grid point $g$ with position $\mathbf r_g$, we write $n_g=n(\mathbf r_g)$ and $n_{\mathrm{ref},g}=n_{\rm ref}(\mathbf r_g)$. The pointwise error $e_g=n_g-n_{\mathrm{ref},g}$ gives the mean absolute error (MAE), the RMSE, and the integrated absolute error per electron $\eta_n$,
\begin{align}
\mathrm{MAE} &= \frac{1}{N_g}\sum_g|e_g|,
\label{eq:density-mae}\\
\mathrm{RMSE} &= \left(\frac{1}{N_g}\sum_g e_g^2\right)^{1/2},
\label{eq:density-rmse}\\
\eta_n &= \frac{\Delta V}{N_e}\sum_g|e_g|.
\label{eq:density-error}
\end{align}
The MAE and RMSE have units of $e\,\text{\AA}^{-3}$, and $\eta_n$ is dimensionless. The charge error is $\delta N=\Delta V\sum_gn_g-N_e$. To evaluate the RMSE in region $\mathcal R$, we divide the sum of squared errors by the number of grid points in that region, $N_{\mathcal R}$, and take the square root. The corresponding relative error is
\begin{equation}
\varepsilon_{2,\mathcal R}=\left(\frac{\sum_{g\in\mathcal R}e_g^2}{\sum_{g\in\mathcal R}n_{\mathrm{ref},g}^2}\right)^{1/2}.
\label{eq:regional-l2}
\end{equation}
Both sums in the relative error run over the same region. The signed integral $\Delta V\sum_{g\in\mathcal R}e_g$ measures regional charge redistribution without imposing a local electron count. The error is undefined for an empty region.

We partition the cell by the minimum periodic distance $d_n$ to a nucleus. In boron systems with vacancies, this distance uses only nuclei of the remaining atoms. The near-nuclear region has $d_n\le0.75$ \AA, the surrounding shell has $0.75<d_n\le1.5$ \AA, and the interstitial region has $d_n>1.5$ \AA. A second partition uses the distance $d_v$ to the nearest removed site, with regions $d_v\le1.5$ \AA, $1.5<d_v\le3$ \AA, and $d_v>3$ \AA. We evaluate the partitions independently, using the full cell geometry for the periodic distances, including for the skew boron cell. Each grid point counts once within a region. The analysis radii are independent of the pseudopotential core radii.

For pristine B, MALA lowers the initial near-nuclear RMSE from $0.091574$ to $0.007234\,e\,\text{\AA}^{-3}$ and the interstitial RMSE from $0.048927$ to $0.002057\,e\,\text{\AA}^{-3}$ (Table~\ref{tab:initial-regions}). For a relative comparison across configurations, we normalize by the norm of the reference density within the interstitial region. The corresponding initial relative error rises from 3.04\% in pristine B to 20.93--22.98\% with four vacancies.

In boron systems with vacancies, the first MALA output has a larger RMSE than ATOMIC near the nuclei and in the surrounding shells. The interstitial RMSE remains smaller with MALA (Table~\ref{tab:first-regions}). Within 1.5 \AA\ of a vacancy site, the relative error rises to 34.90--72.51\% for MALA, compared with 7.96--17.83\% for ATOMIC (Table~\ref{tab:vacancy-regions}). The MALA advantage also disappears in the shell from 1.5 to 3 \AA. Beyond 3 \AA, the MALA output retains the smaller error in all nine cases.

\begin{table*}[tb]
\caption{Initial density RMSE by distance to the nearest nucleus, in $10^{-3}\,e\,\text{\AA}^{-3}$. M and A denote MALA and ATOMIC, both compared with the converged ATOMIC-initialized QE valence density of each configuration. The minimum periodic distance $d_n$ to any nucleus in the cell defines three disjoint regions ($d_n\le0.75$ \AA, $0.75<d_n\le1.5$ \AA, and $d_n>1.5$ \AA). We divide the sum of squared errors by the grid-point count in each region and take the square root. The full periodic geometry and union masks avoid double counting. We evaluate the vacancy-centered partition separately. Bold marks the smaller value in each pair and both values when they tie at the displayed precision.}
\label{tab:initial-regions}
\small
\setlength{\tabcolsep}{3.5pt}
\begin{ruledtabular}
\begin{tabular}{lrrrrrr}
& \multicolumn{2}{c}{Near nuclei} & \multicolumn{2}{c}{Surrounding shell} & \multicolumn{2}{c}{Interstitial} \\
Case & M & A & M & A & M & A \\
\hline
pristine Al298 & \textbf{1.466} & 35.551 & \textbf{1.380} & 17.684 & \textbf{1.209} & 15.947 \\
pristine Al933 & \textbf{5.169} & 35.120 & \textbf{3.715} & 20.688 & \textbf{2.934} & 12.131 \\
pristine B298 & \textbf{7.234} & 91.574 & \textbf{4.064} & 65.931 & \textbf{2.057} & 48.927 \\
B298, 1 vacancy, set 1 & \textbf{15.658} & 91.144 & \textbf{9.797} & 65.130 & \textbf{8.625} & 47.969 \\
B298, 1 vacancy, set 2 & \textbf{16.244} & 91.047 & \textbf{9.981} & 65.187 & \textbf{7.057} & 47.785 \\
B298, 1 vacancy, set 3 & \textbf{16.710} & 91.134 & \textbf{10.647} & 65.206 & \textbf{8.202} & 47.700 \\
B298, 2 vacancies, set 1 & \textbf{21.046} & 90.665 & \textbf{13.214} & 64.490 & \textbf{11.695} & 46.791 \\
B298, 2 vacancies, set 2 & \textbf{21.033} & 90.862 & \textbf{12.930} & 64.440 & \textbf{10.533} & 46.850 \\
B298, 2 vacancies, set 3 & \textbf{21.240} & 90.380 & \textbf{13.580} & 64.390 & \textbf{12.322} & 46.640 \\
B298, 4 vacancies, set 1 & \textbf{28.014} & 89.705 & \textbf{17.591} & 62.888 & \textbf{15.737} & 44.851 \\
B298, 4 vacancies, set 2 & \textbf{27.290} & 89.631 & \textbf{17.307} & 62.898 & \textbf{14.368} & 44.938 \\
B298, 4 vacancies, set 3 & \textbf{27.965} & 89.876 & \textbf{17.489} & 62.974 & \textbf{14.381} & 44.935 \\
\end{tabular}
\end{ruledtabular}
\end{table*}

\begin{table*}[tb]
\caption{Density RMSE of the first output before mixing, in $10^{-3}\,e\,\text{\AA}^{-3}$ and at finite Davidson accuracy. M and A denote MALA and ATOMIC, both compared with the converged ATOMIC-initialized QE valence density of each configuration. The minimum periodic distance $d_n$ to any nucleus in the cell defines three disjoint regions ($d_n\le0.75$ \AA, $0.75<d_n\le1.5$ \AA, and $d_n>1.5$ \AA). We divide the sum of squared errors by the grid-point count in each region and take the square root. The full periodic geometry and union masks avoid double counting. We evaluate the vacancy-centered partition separately. Bold marks the smaller value in each pair and both values when they tie at the displayed precision.}
\label{tab:first-regions}
\small
\setlength{\tabcolsep}{3.5pt}
\begin{ruledtabular}
\begin{tabular}{lrrrrrr}
& \multicolumn{2}{c}{Near nuclei} & \multicolumn{2}{c}{Surrounding shell} & \multicolumn{2}{c}{Interstitial} \\
Case & M & A & M & A & M & A \\
\hline
pristine Al298 & \textbf{1.365} & 4.007 & \textbf{1.829} & 4.440 & \textbf{1.455} & 3.361 \\
pristine Al933 & 8.615 & \textbf{6.631} & 11.329 & \textbf{8.723} & 7.962 & \textbf{5.585} \\
pristine B298 & \textbf{8.220} & 52.153 & \textbf{3.984} & 28.472 & \textbf{0.940} & 12.896 \\
B298, 1 vacancy, set 1 & 75.703 & \textbf{52.527} & 40.303 & \textbf{28.323} & \textbf{9.823} & 12.798 \\
B298, 1 vacancy, set 2 & 68.510 & \textbf{53.798} & 36.863 & \textbf{28.978} & \textbf{10.702} & 13.331 \\
B298, 1 vacancy, set 3 & 75.476 & \textbf{53.170} & 40.814 & \textbf{28.823} & \textbf{12.193} & 13.425 \\
B298, 2 vacancies, set 1 & 70.416 & \textbf{54.093} & 37.046 & \textbf{28.949} & \textbf{8.978} & 13.222 \\
B298, 2 vacancies, set 2 & 60.163 & \textbf{55.227} & 31.282 & \textbf{29.340} & \textbf{8.738} & 13.499 \\
B298, 2 vacancies, set 3 & 73.318 & \textbf{55.294} & 38.952 & \textbf{29.326} & \textbf{10.465} & 13.310 \\
B298, 4 vacancies, set 1 & 76.868 & \textbf{55.999} & 40.118 & \textbf{29.252} & \textbf{10.892} & 13.402 \\
B298, 4 vacancies, set 2 & 77.062 & \textbf{56.661} & 39.745 & \textbf{29.565} & \textbf{10.996} & 13.586 \\
B298, 4 vacancies, set 3 & 73.270 & \textbf{56.534} & 37.448 & \textbf{29.355} & \textbf{9.824} & 13.433 \\
\end{tabular}
\end{ruledtabular}
\end{table*}

\begin{table*}[tb]
\caption{Relative $L^2$ density errors around boron vacancies, in percent and listed as MALA/ATOMIC. We take the converged ATOMIC-initialized QE valence density of the same configuration as the reference and restrict both norms to the same grid points. The minimum periodic distance $d_v$ to a removed site defines the center ($d_v\le1.5$ \AA), the shell ($1.5<d_v\le3$ \AA), and the far region ($d_v>3$ \AA). The radii are independent of the pseudopotential cores. Points in overlapping vacancy spheres count once, and the partition can overlap the regions defined by distance to the nearest nucleus. First denotes the output before the first mixing step, at finite Davidson accuracy. Bold marks the smaller value in each pair and both values when they tie at the displayed precision.}
\label{tab:vacancy-regions}
\footnotesize
\setlength{\tabcolsep}{3.5pt}
\begin{ruledtabular}
\begin{tabular}{lrrrrrr}
& \multicolumn{3}{c}{Initial (M/A)} & \multicolumn{3}{c}{First response (M/A)} \\
Case & Center & Shell & Far & Center & Shell & Far \\
\hline
B298, 1 vacancy, set 1 & 16.64/\textbf{13.60} & \textbf{5.06}/14.69 & \textbf{1.24}/14.89 & 59.94/\textbf{7.96} & 26.23/\textbf{7.31} & \textbf{4.72}/7.30 \\
B298, 1 vacancy, set 2 & 20.55/\textbf{16.10} & \textbf{4.48}/14.77 & \textbf{1.22}/14.86 & 67.38/\textbf{17.83} & 21.83/\textbf{8.18} & \textbf{4.34}/7.28 \\
B298, 1 vacancy, set 3 & 22.01/\textbf{16.47} & \textbf{4.72}/14.94 & \textbf{1.25}/14.85 & 72.51/\textbf{16.55} & 25.27/\textbf{7.83} & \textbf{4.52}/7.27 \\
B298, 2 vacancies, set 1 & 18.25/\textbf{15.26} & \textbf{4.77}/14.79 & \textbf{1.48}/14.79 & 46.16/\textbf{14.05} & 16.07/\textbf{7.64} & \textbf{4.70}/7.35 \\
B298, 2 vacancies, set 2 & 18.10/\textbf{15.13} & \textbf{4.56}/14.65 & \textbf{1.48}/14.84 & 42.50/\textbf{15.52} & 12.85/\textbf{7.85} & \textbf{3.96}/7.41 \\
B298, 2 vacancies, set 3 & 19.17/\textbf{14.51} & \textbf{4.79}/14.67 & \textbf{1.53}/14.79 & 51.78/\textbf{16.38} & 16.48/\textbf{8.08} & \textbf{4.76}/7.34 \\
B298, 4 vacancies, set 1 & 17.00/\textbf{14.43} & \textbf{4.68}/14.68 & \textbf{2.13}/14.63 & 37.63/\textbf{13.54} & 11.13/\textbf{7.64} & \textbf{6.86}/7.50 \\
B298, 4 vacancies, set 2 & 16.46/\textbf{14.62} & \textbf{4.50}/14.74 & \textbf{2.08}/14.58 & 36.36/\textbf{13.88} & 10.76/\textbf{7.88} & \textbf{7.28}/7.44 \\
B298, 4 vacancies, set 3 & 17.02/\textbf{14.60} & \textbf{4.56}/14.62 & \textbf{2.14}/14.70 & 34.90/\textbf{13.54} & 10.76/\textbf{7.81} & \textbf{6.30}/7.46 \\
\end{tabular}
\end{ruledtabular}
\end{table*}

\subsection{Reciprocal-space errors}

We define the Fourier coefficient $\widehat e(\mathbf G)$ of the density error by
\begin{equation}
\widehat e(\mathbf G)=\frac{1}{N_g}\sum_g e_g\,e^{-i\mathbf G\cdot\mathbf r_g}.
\label{eq:density-error-fourier}
\end{equation}
For the $\Gamma$ half-sphere stored by QE, the multiplicity is $m_{\mathbf G}=2$ for nonzero conjugate pairs and $m_{\mathbf 0}=1$ for the zero mode. With these weights, Parseval's identity gives
\begin{equation}
\frac{1}{N_g}\sum_g e_g^2=\sum_{\mathbf G\,\mathrm{stored}}m_{\mathbf G}|\widehat e(\mathbf G)|^2.
\label{eq:parseval}
\end{equation}
We verify the Parseval sums against the real-space errors. The cell electron counts agree to below $10^{-10}\,e$. We retain the zero coefficient in real-space and charge measures, and omit it from the Fourier and Hartree energy plots to isolate the errors at nonzero wave vectors.

To resolve the error by wave-vector magnitude, we divide the native reciprocal sphere into 80 equal-width radial shells. The density-error power in shell $b$ is
\begin{equation}
D_b=\sum_{\substack{\mathbf G\in b\,\mathrm{stored}\\\mathbf G\ne0}}m_{\mathbf G}|\widehat e(\mathbf G)|^2.
\label{eq:shell-power}
\end{equation}
The plots give each shell contribution as $\sqrt{D_b}$ in $e\,\text{\AA}^{-3}$, without averaging over modes. We normalize each cumulative curve by its own $\sum_bD_b$ to obtain the distribution of error power. We show the amplification $\sqrt{D_b^{\rm first}/D_b^{\rm initial}}$, where the superscripts denote the first unmixed output and the starting density, respectively. We report the ratio for shells whose initial power exceeds $10^{-12}$ of the total initial power.

We apply the Hartree kernel to obtain the Fourier coefficient $\bar{v}_{\mathrm H}(\mathbf G)$ of the potential difference and take its RMS,
\begin{align}
\delta v_{\mathrm H}(\mathbf G)&=\frac{4\pi}{G^2}\widehat e(\mathbf G),\quad \mathbf G\ne0,\\
H&=\left(\sum_{\substack{\mathbf G\,\mathrm{stored}\\\mathbf G\ne0}}m_{\mathbf G}|\delta v_{\mathrm H}(\mathbf G)|^2\right)^{1/2},
\label{eq:hartree-rms}
\end{align}
where we set $\delta v_{\mathrm H}(\mathbf 0)=0$. The density and reciprocal coordinates use atomic units, and we convert $\bar{v}_{\mathrm H}$ using $1$ hartree $=27.211386245988$ eV. The resulting $\bar{v}_{\mathrm H}$ is the RMS error relative to the zero-mean potential of $n_{\rm ref}$. It is distinct from both QE's SCF estimate, which compares the input and output of the same iteration, and a total-energy difference. The dependence on the converged reference makes $\bar{v}_{\mathrm H}$ a diagnostic.

\begin{figure*}[!t]
\centering
\includegraphics[width=0.84\textwidth]{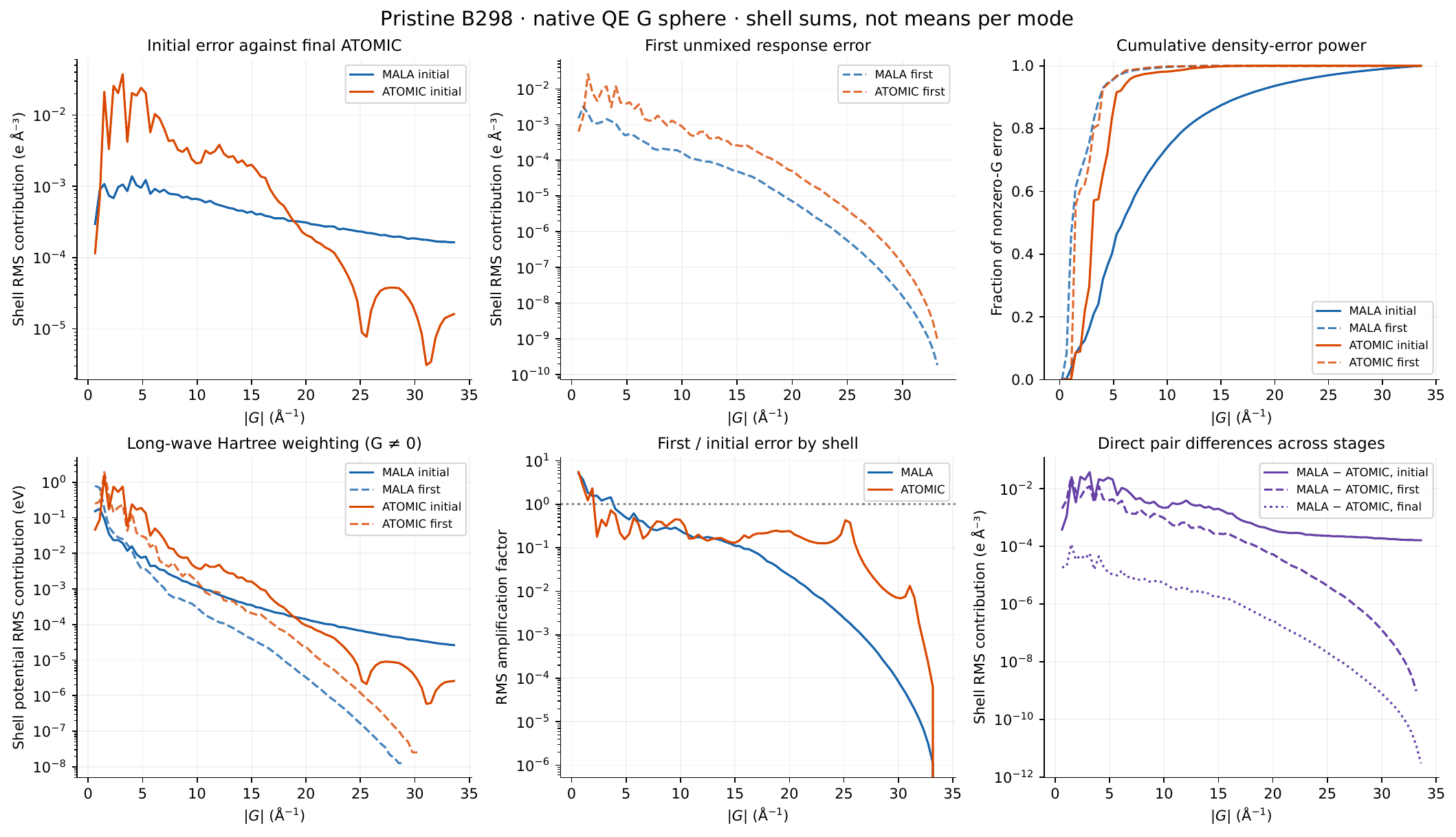}
\caption{Fourier density errors for pristine B298. Top: shell-resolved MALA/ATOMIC errors for the initial (left) and first unmixed (middle) densities relative to the converged ATOMIC-initialized QE valence density $n_{\rm ref}$, with normalized cumulative error power (right). Bottom: Hartree-weighted potential-error contributions (left), first-to-initial shell-error ratios (middle), and direct MALA--ATOMIC differences at the initial, first, and final stages (right). Density spectra sum nonzero Fourier modes in 80 radial shells, including conjugate-pair multiplicities; shell contributions are in $e\,\text{\AA}^{-3}$, potential contributions in eV, and $|\mathbf G|$ in \AA$^{-1}$. First responses precede mixing and retain finite Davidson accuracy. Hartree-weighted errors are distinct from QE's printed SCF residual.}
\label{fig:spectra-b298}
\vspace{4pt}
\centering
\includegraphics[width=0.84\textwidth]{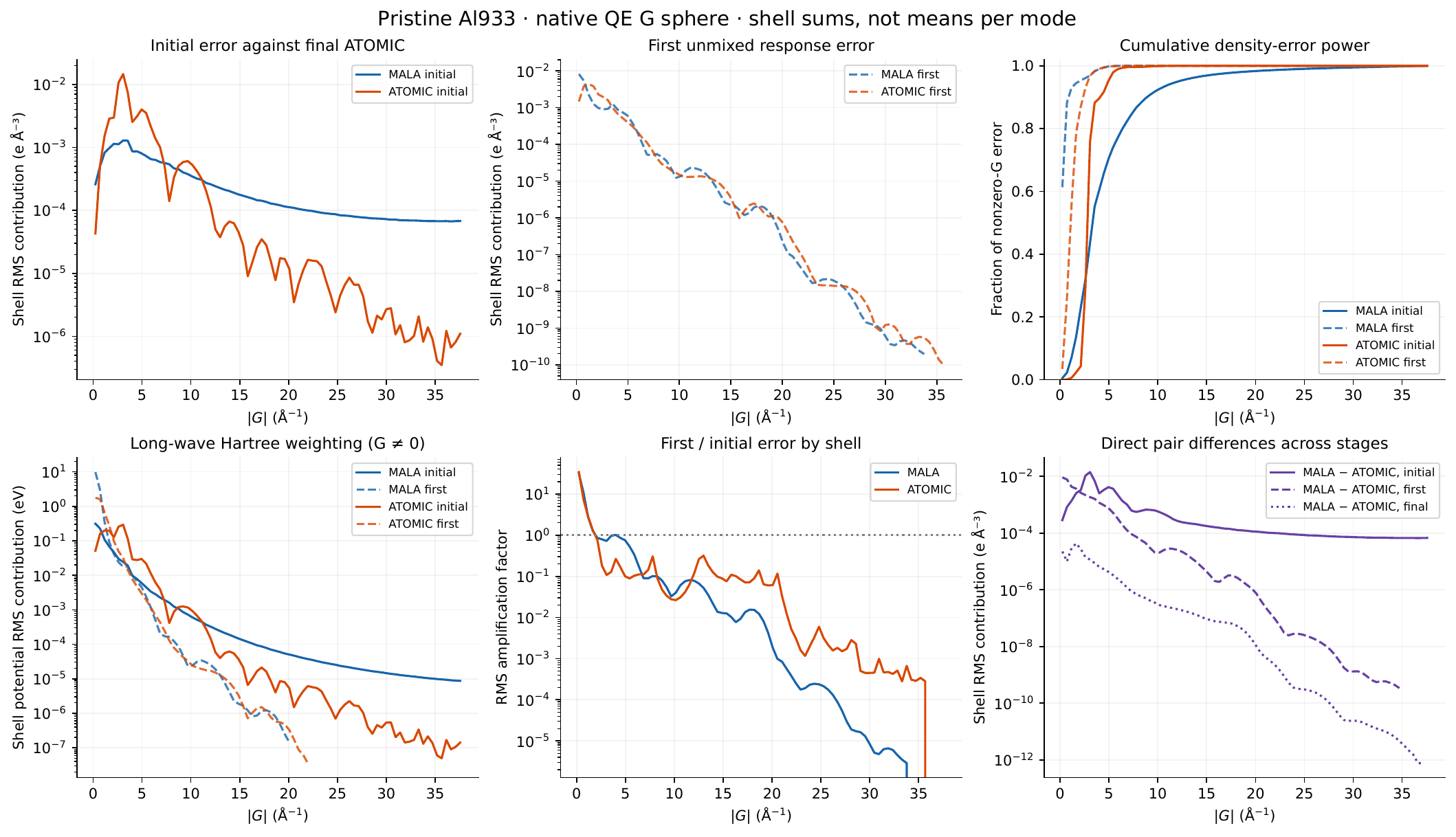}
\caption{Fourier density errors for pristine Al933. Top: shell-resolved MALA/ATOMIC errors for the initial (left) and first unmixed (middle) densities relative to the converged ATOMIC-initialized QE valence density $n_{\rm ref}$, with normalized cumulative error power (right). Bottom: Hartree-weighted potential-error contributions (left), first-to-initial shell-error ratios (middle), and direct MALA--ATOMIC differences at the initial, first, and final stages (right). Density spectra sum nonzero Fourier modes in 80 radial shells, including conjugate-pair multiplicities; shell contributions are in $e\,\text{\AA}^{-3}$, potential contributions in eV, and $|\mathbf G|$ in \AA$^{-1}$. First responses precede mixing and retain finite Davidson accuracy. Hartree-weighted errors are distinct from QE's printed SCF residual.}
\label{fig:spectra-al933}
\end{figure*}

\begin{figure*}[!t]
\centering
\includegraphics[width=\textwidth]{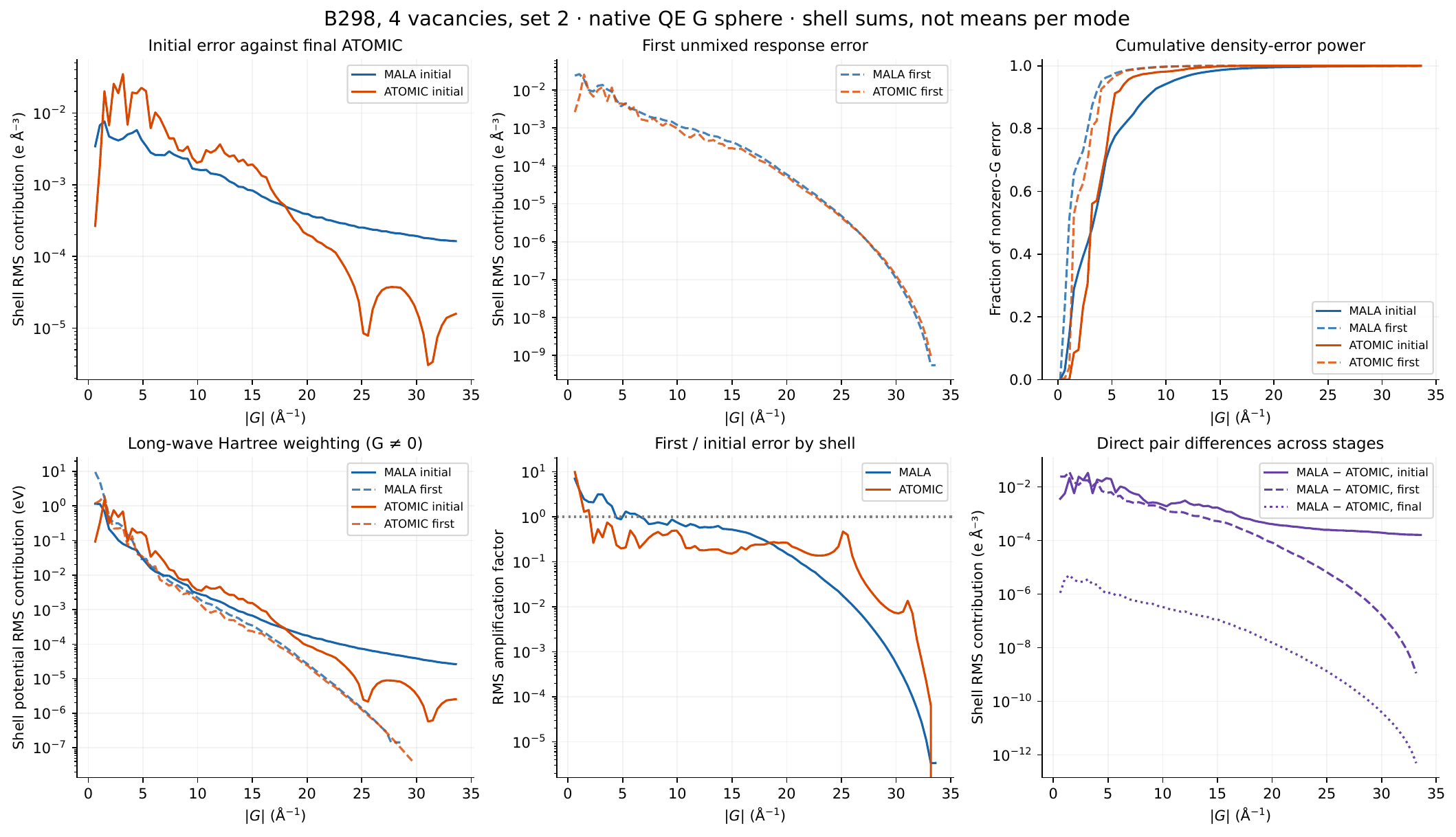}
\caption{Fourier density errors for B298 (4 vacancies, set 2). Top: shell-resolved MALA/ATOMIC errors for the initial (left) and first unmixed (middle) densities relative to the converged ATOMIC-initialized QE valence density $n_{\rm ref}$, with normalized cumulative error power (right). Bottom: Hartree-weighted potential-error contributions (left), first-to-initial shell-error ratios (middle), and direct MALA--ATOMIC differences at the initial, first, and final stages (right). Density spectra sum nonzero Fourier modes in 80 radial shells, including conjugate-pair multiplicities; shell contributions are in $e\,\text{\AA}^{-3}$, potential contributions in eV, and $|\mathbf G|$ in \AA$^{-1}$. First responses precede mixing and retain finite Davidson accuracy. Hartree-weighted errors are distinct from QE's printed SCF residual.}
\label{fig:spectra-v4}
\end{figure*}

Figures~\ref{fig:spectra-b298}--\ref{fig:spectra-v4} resolve the density errors by wavelength in pristine B298, Al933, and B298 with four vacancies. They compare the initial MALA density with the converged reference and give the MALA--ATOMIC differences at corresponding stages. In Al933, the first MALA step raises the Hartree-potential-error RMS from 0.413 to 10.309 eV, compared with 2.503 eV after the ATOMIC step. In Al298, MALA retains the smaller global density error after the first step and saves four iterations, even though its first Hartree error is 1.634 eV compared with 1.244 eV for ATOMIC. Pristine B298 is the sole case in which the first MALA output has the smaller Hartree error.

The growth of the error at small $|\mathbf G|$ and the $G^{-2}$ Hartree weighting are consistent with sensitivity to long-wavelength potential errors. Finite Davidson accuracy and changes in occupations and in the exchange-correlation potential also enter the first response. The diagnostics therefore leave the mechanism undetermined. Damping and mixing history shape the inputs to later iterations.

We write $\mathbf G=\sum_{i=1}^{3}h_i\mathbf b_i$, where $\mathbf b_i$ are the reciprocal-lattice basis vectors and $h_i$ are integer Miller indices. The maps show the $h_3=0$ slice for $|\mathbf G|\le6\,\text{\AA}^{-1}$ in Cartesian reciprocal coordinates. Each map gives $|\Delta n_{\mathbf G}|$ in $e\,\text{\AA}^{-3}$, where $\Delta n_{\mathbf G}$ is the difference between the Fourier coefficients of the two densities being compared. The six panels for a given configuration share a logarithmic scale, which can vary across configurations. We mask the zero mode. To display the remaining modes, we assign the lowest color to values below $\max(10^{-6}v_{\max},10^{-11}\,e\,\text{\AA}^{-3})$, where $v_{\max}$ is the largest value across the six maps. The numerical metrics use the underlying values.

\begin{figure*}[!t]
\centering
\includegraphics[width=0.84\textwidth]{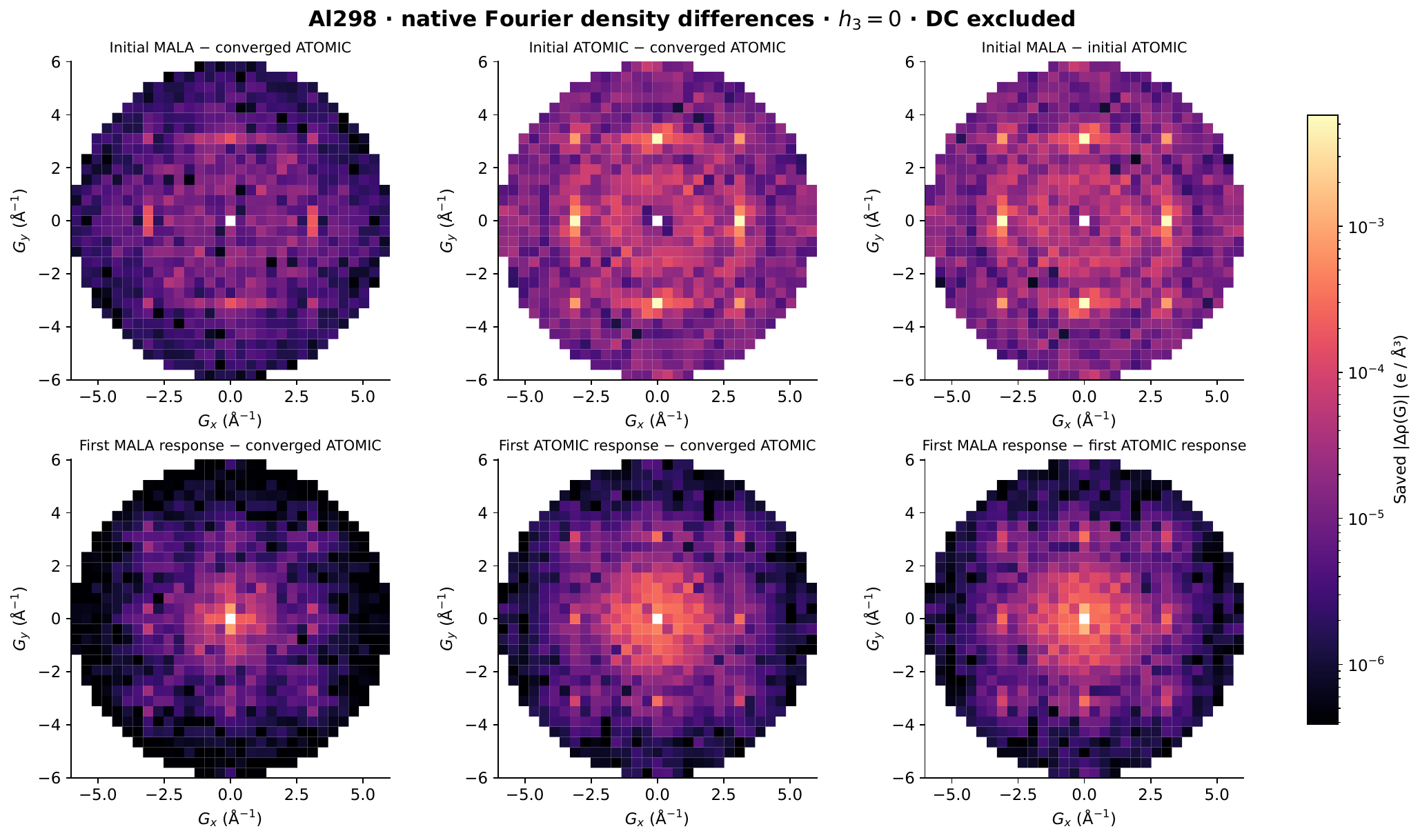}
\caption{Fourier density difference maps for pristine Al298. The magnitudes $|\Delta n_{\mathbf G}|$ in $e\,\text{\AA}^{-3}$ are shown for the $h_3=0$ reciprocal plane with $|\mathbf G|\le6\,\text{\AA}^{-1}$. Top: initial MALA--reference, ATOMIC--reference, and MALA--ATOMIC differences (left to right). Bottom: the corresponding first unmixed differences at finite Davidson accuracy. The reference $n_{\rm ref}$ is the converged ATOMIC-initialized QE valence density. The plot symbol $\rho$ denotes $n$. Cartesian reciprocal axes are in \AA$^{-1}$, and $\mathbf G=0$ is masked. The six panels share a logarithmic color scale with floor $\max(10^{-6}v_{\max},10^{-11}\,e\,\text{\AA}^{-3})$, where $v_{\max}$ is their largest amplitude. The floor affects display only.}
\label{fig:maps-al298}
\vspace{4pt}
\centering
\includegraphics[width=0.84\textwidth]{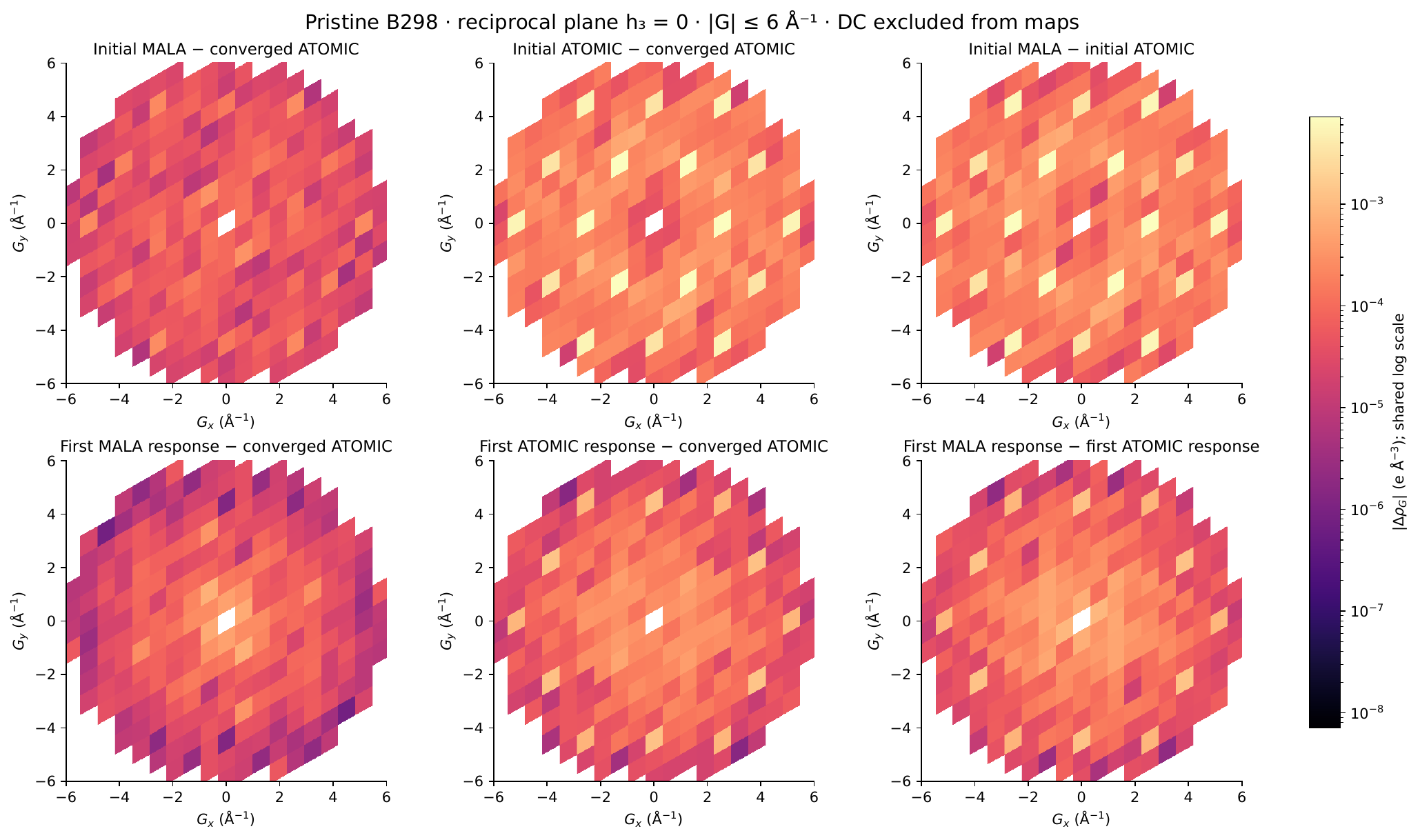}
\caption{Fourier density difference maps for pristine B298. The magnitudes $|\Delta n_{\mathbf G}|$ in $e\,\text{\AA}^{-3}$ are shown for the $h_3=0$ reciprocal plane with $|\mathbf G|\le6\,\text{\AA}^{-1}$. Top: initial MALA--reference, ATOMIC--reference, and MALA--ATOMIC differences (left to right). Bottom: the corresponding first unmixed differences at finite Davidson accuracy. The reference $n_{\rm ref}$ is the converged ATOMIC-initialized QE valence density. Cartesian reciprocal axes are in \AA$^{-1}$, and $\mathbf G=0$ is masked. The six panels share a logarithmic color scale with floor $\max(10^{-6}v_{\max},10^{-11}\,e\,\text{\AA}^{-3})$, where $v_{\max}$ is their largest amplitude. The floor affects display only.}
\label{fig:maps-b298}
\end{figure*}

\begin{figure*}[!t]
\centering
\includegraphics[width=0.84\textwidth]{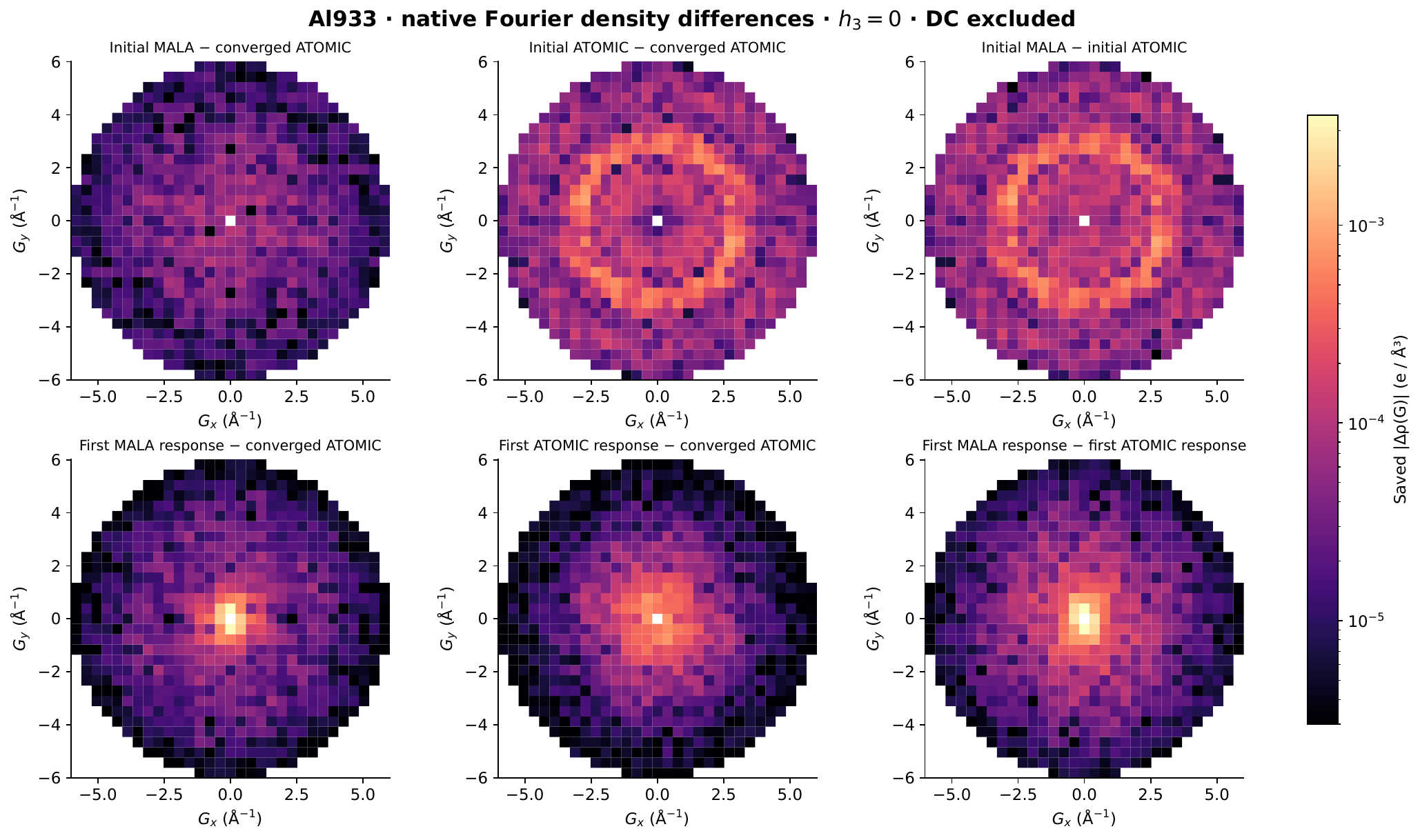}
\caption{Fourier density difference maps for pristine Al933. The magnitudes $|\Delta n_{\mathbf G}|$ in $e\,\text{\AA}^{-3}$ are shown for the $h_3=0$ reciprocal plane with $|\mathbf G|\le6\,\text{\AA}^{-1}$. Top: initial MALA--reference, ATOMIC--reference, and MALA--ATOMIC differences (left to right). Bottom: the corresponding first unmixed differences at finite Davidson accuracy. The reference $n_{\rm ref}$ is the converged ATOMIC-initialized QE valence density. The plot symbol $\rho$ denotes $n$. Cartesian reciprocal axes are in \AA$^{-1}$, and $\mathbf G=0$ is masked. The six panels share a logarithmic color scale with floor $\max(10^{-6}v_{\max},10^{-11}\,e\,\text{\AA}^{-3})$, where $v_{\max}$ is their largest amplitude. The floor affects display only.}
\label{fig:maps-al933}
\vspace{4pt}
\centering
\includegraphics[width=0.84\textwidth]{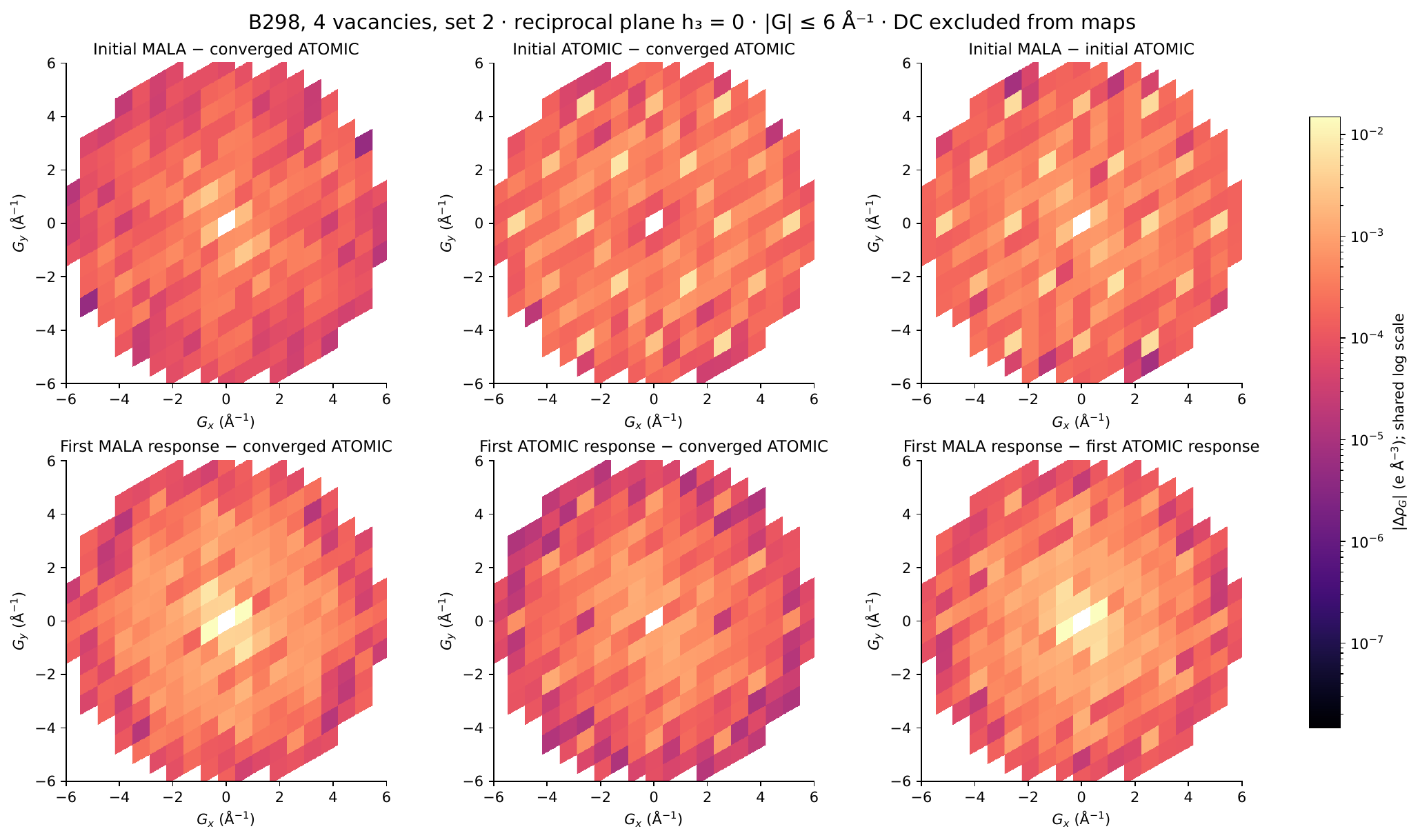}
\caption{Fourier density difference maps for B298 (4 vacancies, set 2). The magnitudes $|\Delta n_{\mathbf G}|$ in $e\,\text{\AA}^{-3}$ are shown for the $h_3=0$ reciprocal plane with $|\mathbf G|\le6\,\text{\AA}^{-1}$. Top: initial MALA--reference, ATOMIC--reference, and MALA--ATOMIC differences (left to right). Bottom: the corresponding first unmixed differences at finite Davidson accuracy. The reference $n_{\rm ref}$ is the converged ATOMIC-initialized QE valence density. Cartesian reciprocal axes are in \AA$^{-1}$, and $\mathbf G=0$ is masked. The six panels share a logarithmic color scale with floor $\max(10^{-6}v_{\max},10^{-11}\,e\,\text{\AA}^{-3})$, where $v_{\max}$ is their largest amplitude. The floor affects display only.}
\label{fig:maps-v4}
\end{figure*}

The radial sums combine all vectors of similar magnitude, whereas the maps in Figs.~\ref{fig:maps-al298}--\ref{fig:maps-v4} show the directional structure of the initial and first-output differences within one reciprocal plane. The three-dimensional error norm uses the full reciprocal sphere. Each shell combines contributions from across the real-space cell, including the regions near nuclei and around vacancies.

\clearpage
\section{Earlier CP2K density-transfer controls}
\label{app:cp2k-controls}

Earlier tests in CP2K motivated the matched QE comparisons. A density converged in one representation can follow a different path to convergence when transferred to another. CP2K's GPW method expands the orbitals in Gaussians and represents the density on an auxiliary plane-wave grid. QE expands the orbitals in plane waves~\cite{VandeVondele2005,Kuhne2020}. In CP2K, the 800 Ry cutoff applies to the density grid, while the QE wave-function cutoff applies to the orbitals. Equal grid dimensions alone leave the equivalence of the representations undetermined. The controls track convergence after transferring the densities under the settings used by CP2K.

\subsection{Hamiltonians, initialization, and stopping rules}

One group of CP2K calculations uses PBE with Goedecker--Teter--Hutter (GTH) pseudopotentials, three valence electrons per atom, and the \texttt{DZVP-MOLOPT-SR-GTH} basis~\cite{Perdew1996,Goedecker1996,Krack2005,VandeVondele2007}. These runs use \texttt{CUTOFF=800 Ry} and \texttt{REL\_CUTOFF=60 Ry}, with a principal cutoff of 8000 Ry for the historical pristine-boron pair. All use $\Gamma$ sampling, periodic boundaries in three dimensions, diagonalization, Fermi--Dirac occupations, and \texttt{BROYDEN\_MIXING} with \texttt{ALPHA=0.10} and \texttt{NBUFFER=8}.

The electronic temperatures are 300 K for Al298, 931.5366317709 K for Al933, and 298 K for boron. The Al and B inputs request 256 and 64 additional orbitals, respectively. Each MALA/ATOMIC pair shares the geometry and numerical settings, and the production runs use \texttt{MAX\_SCF=50}. The GTH/PBE Al298 own-density control allows eight steps and converges in four.

For the CP2K inner SCF cycle, we use \texttt{EPS\_SCF} of $10^{-5}$ in the pristine cases and $10^{-7}$ in the vacancy cases. QE's \texttt{conv\_thr} limits the estimated SCF energy error in Ry/cell, so the same numerical exponent corresponds to different physical stopping criteria in the two codes. We chose the tighter vacancy setting to obtain stable reference densities and energies. Observable convergence at that threshold remains untested.

Control $R_{\rm C}$ initializes CP2K with a valence density converged for the same CP2K Hamiltonian. For $R_{\rm Q}$, we integrate the original dense-mesh QE-reference LDOS to obtain the scalar starting density. Both controls transfer the density alone and start without a native electronic restart. For Al298 GTH/PBE at 800 Ry, $R_{\rm C}$ starts from the MALA-initialized endpoint. For Al933 GTH/PBE and the Al298 control with an NC pseudopotential in UPF and the PBEsol functional (NC-UPF/PBEsol), $R_{\rm C}$ starts from the ATOMIC CP2K endpoint.

For Al at 800 Ry, CP2K projects the external $200^3$ source cube onto its finest $288^3$ grid. The native $R_{\rm C}$ cubes already use that grid. Import checks document the grid selection and the corrections to voxel-vector precision. The CP2K UPF importer applies a near-unity charge normalization independently of the QE converter.

Let $N_{\rm A}$, $N_{\rm M}$, $N_{\rm C}$, and $N_{\rm Q}$ count the completed CP2K iterations for ATOMIC, MALA, $R_{\rm C}$, and $R_{\rm Q}$, respectively. We use these counts to define the iteration fractions
\begin{align}
 \eta_{N,\rm C}&=\frac{N_{\rm A}-N_{\rm M}}{N_{\rm A}-N_{\rm C}},\qquad
 \eta_{N,\rm Q}=\frac{N_{\rm A}-N_{\rm M}}{N_{\rm A}-N_{\rm Q}}.
 \label{eq:cp2k-fractions}
\end{align}
The denominator of $\eta_{N,\rm C}$ uses the observed count for $R_{\rm C}$, and the denominator of $\eta_{N,\rm Q}$ uses the observed count for $R_{\rm Q}$. Both controls supply empirical counts and leave the minimum iteration count and theoretical upper bound undetermined. A value of $\eta_{N,\rm Q}=100\%$ means that MALA and the transferred QE-reference density take the same number of steps.

\begin{table*}[tb]
\caption{Pristine CP2K controls at \texttt{EPS\_SCF}$=10^{-5}$. The cutoffs apply to the auxiliary density grid and are given in Ry, with \texttt{REL\_CUTOFF}$=60$ Ry. The original logs give the number of iterations to convergence. $M$, $A$, $C$, and $Q$ denote MALA, ATOMIC, reinjection of a CP2K density $R_{\rm C}$, and transfer of the density from the original dense-mesh QE-reference LDOS $R_{\rm Q}$, respectively. We express the ratios in Eq.~\ref{eq:cp2k-fractions} in percent. Bold compares MALA with ATOMIC and marks both at an exact tie. The signed difference per atom, $\Delta F/N_{\rm at}$, is given in meV/atom, with $\Delta F=F_{\rm M}-F_{\rm A}$. Here $F_{\rm M}$ and $F_{\rm A}$ are the converged CP2K energies for MALA and ATOMIC, respectively. These printed finite-temperature energies include the electronic entropy contribution. $N_{\rm at}=256$ for aluminum and 144 for pristine boron. A dash marks an unavailable calculation or ratio. The 8000-Ry B $R_{\rm C}$ control used the same numerical settings in a later allocation and provides an iteration count.}
\label{tab:cp2k-pristine}
\small
\setlength{\tabcolsep}{3.5pt}
\begin{ruledtabular}
\begin{tabular}{lr rrrr rr r}
Case and Hamiltonian & Cutoff & $N_{\rm M}$ & $N_{\rm A}$ & $N_{\rm C}$ & $N_{\rm Q}$ & $\eta_{N,\rm C}$ (\%) & $\eta_{N,\rm Q}$ (\%) & $\Delta F/N_{\rm at}$ \\
\hline
Al298, GTH/PBE & 800 & \textbf{15} & 17 & 4 & 14 & 15.38 & 66.67 & $+0.119500$ \\
Al933, GTH/PBE & 800 & \textbf{17} & \textbf{17} & 4 & 15 & 0.00 & 0.00 & $+0.046701$ \\
B298, GTH/PBE & 800 & -- & 16 & -- & 12 & -- & -- & -- \\
B298, GTH/PBE (historical) & 8000 & \textbf{14} & 16 & 2 & -- & 14.29 & -- & $-0.117583$ \\
Al298, NC-UPF/PBEsol & 800 & \textbf{14} & 17 & 4 & 14 & 23.08 & 100.00 & $+0.110873$ \\
\end{tabular}
\end{ruledtabular}
\end{table*}

\subsection{Pristine controls and Hamiltonian sensitivity}

\begin{figure*}[!t]
\centering
\includegraphics[width=\textwidth]{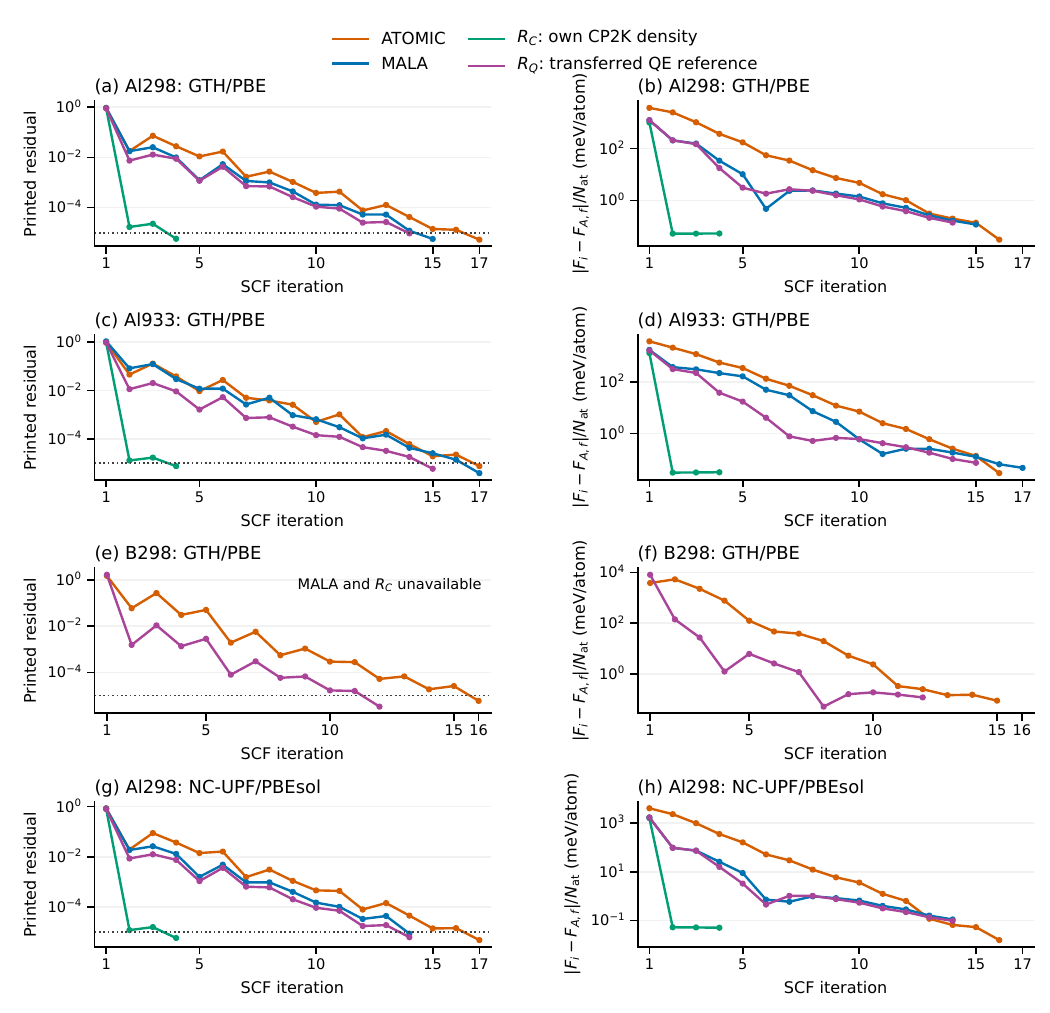}
\caption{Pristine 800-Ry CP2K trajectories for GTH/PBE Al298, Al933, and B298, followed by NC-UPF/PBEsol Al298. The left column gives the printed CP2K residual in its native normalization, with the dotted line at \texttt{EPS\_SCF}$=10^{-5}$. QE's estimated energy error and the Hartree-weighted Fourier diagnostic are distinct quantities. The right column gives $|F^{(\ell)}-F_{\rm A}|/N_{\rm at}$ in meV/atom, referenced to the converged ATOMIC energy for the same geometry and Hamiltonian. The axes use $F_i$ for the finite-temperature energy $F^{(\ell)}$ at iteration $\ell=i$, and $F_{\rm A,f}$ for $F_{\rm A}$. The printed energies include the electronic entropy contribution. CP2K gives the extrapolated zero-temperature energy separately. The conversion uses $1\,{\rm Ha}=27.211386245988$ eV, with $N_{\rm at}=256$ for Al and 144 for B. The logarithmic axis omits exact zeros, including the last ATOMIC point, and shows the finite nonzero energy offsets. Each residual curve ends at convergence. $R_{\rm C}$ reinjects the converged CP2K density. $R_{\rm Q}$ transfers the density obtained by integrating the original dense-mesh QE-reference LDOS. For B298, MALA and $R_{\rm C}$ are unavailable at 800 Ry. Table~\ref{tab:cp2k-pristine} gives the separate 8000-Ry result.}
\label{fig:cp2k-pristine}
\end{figure*}

With GTH/PBE, MALA saves two steps for Al298 and ties ATOMIC for Al933 (Table~\ref{tab:cp2k-pristine}). The own-density controls take four steps in both cells, and the transferred QE-reference densities take 14 and 15 steps, respectively. For pristine boron, ATOMIC and $R_{\rm Q}$ are available at 800 Ry. The MALA and ATOMIC counts of 14 and 16 come from the historical 8000-Ry comparison. The different cutoffs define separate comparisons, and Fig.~\ref{fig:cp2k-pristine} shows the available 800-Ry trajectories.

The iteration savings are smaller or absent in some CP2K tests. In liquid aluminum, for example, MALA saves iterations in QE and ties the ATOMIC guess in CP2K. The comparison suggests that the gains from a better starting density depend on the code and the calculation settings.

For the additional Al298 comparison, we use PBEsol~\cite{Perdew2008PBEsol} with the original NC \texttt{Al.sr-pbesol.upf} pseudopotential, including its nonlinear core correction. The Gaussian basis and 800/60-Ry grid settings are the same as before. MALA takes 14 steps, ATOMIC 17, $R_{\rm C}$ four, and $R_{\rm Q}$ 14. The saving is two steps in GTH/PBE and three in NC-UPF/PBEsol. The executable also changed, and the newer executable lacks a GTH/PBE control in this comparison. This leaves the individual effects of the pseudopotential and functional undetermined. The transferred QE-reference density still takes 14 steps. The basis, source k-point sampling, representation, solver, and finite tolerance all remain possible contributors.

Both UPF production runs reach \texttt{EPS\_SCF}$=10^{-5}$. Their final printed energies differ by $0.001043070908$ Ha per cell, or $0.11087$ meV/atom, slightly above the earlier $0.001$-Ha flag. Table~\ref{tab:cp2k-pristine} lists the signed differences at finite tolerance. The plotted $F$ includes electronic entropy. CP2K prints an extrapolated $T_e\to0$ energy separately.

\subsection{Boron vacancies and threshold dependence}

\begin{figure*}[!t]
\centering
\includegraphics[width=\textwidth]{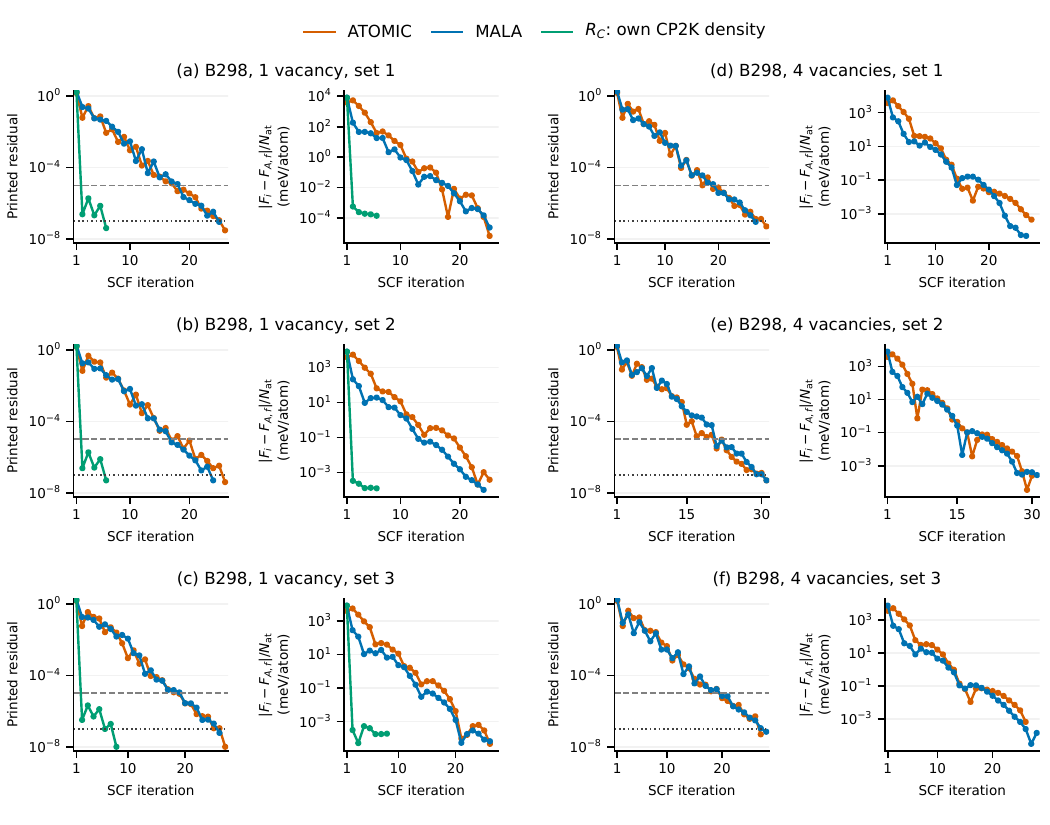}
\caption{CP2K trajectories for six vacancy geometries with GTH/PBE at 800/60 Ry. The left panels give the printed SCF residual up to convergence. The dotted line marks the target \texttt{EPS\_SCF}$=10^{-7}$, and the dashed line marks $10^{-5}$. We read the first crossing of the dashed line from the same trajectory. The right panels give $|F^{(\ell)}-F_{\rm A}|/N_{\rm at}$ in meV/atom, referenced to the converged ATOMIC finite-temperature energy for the same geometry. The axis notation, unit conversion, and omission of exact zero offsets on logarithmic axes follow Fig.~\ref{fig:cp2k-pristine}. $N_{\rm at}=143$ for one vacancy and 140 for four. Reinjection of the converged CP2K density ($R_{\rm C}$) is available for the three single vacancies. The trajectories include savings, ties, and an increase in iterations with MALA.}
\label{fig:cp2k-vacancies}
\end{figure*}

The CP2K vacancy pairs share six geometries with QE, three with one vacancy and three with four (Table~\ref{tab:cp2k-vacancies} and Fig.~\ref{fig:cp2k-vacancies}). These calculations use GTH/PBE, 800/60 Ry, and \texttt{EPS\_SCF}$=10^{-7}$. Matched CP2K baselines for the divacancies are unavailable. For the single vacancies, MALA saves one or two steps, and the own-density controls take six, six, and eight steps. For four vacancies, changing the placement can save two steps with MALA, give a tie, or add one step. The savings thus depend on the configuration, and the selected cells leave a monotonic concentration boundary undetermined.

We locate the first $10^{-5}$ crossing within each trajectory run to $10^{-7}$. At that earlier crossing, MALA takes one extra step for single-vacancy sets 1 and 3 and for four-vacancy set 1. The other three pairs tie. A later crossing at $10^{-5}$ can therefore occur in a MALA trajectory that converges sooner at $10^{-7}$. The pristine and vacancy calculations use different targets, and CP2K and QE use different measures of convergence.

\begin{table*}[tb]
\caption{CP2K boron vacancy calculations with GTH/PBE at 800/60 Ry. $N_{\rm M}$, $N_{\rm A}$, and $N_{\rm C}$ count iterations to \texttt{EPS\_SCF}$=10^{-7}$. We express the reinjection-referenced fraction $\eta_{N,\rm C}$ from Eq.~\ref{eq:cp2k-fractions} in percent. Bold compares MALA and ATOMIC and marks both at an exact tie. The first crossing gives the first iteration below $10^{-5}$ in the same $10^{-7}$ run, in the order MALA/ATOMIC/$R_{\rm C}$. The signed difference per atom, $\Delta F/N_{\rm at}$, is given in $\mu$eV/atom and uses the CP2K energies defined in Table~\ref{tab:cp2k-pristine}. $N_{\rm at}=143$ for one vacancy and 140 for four. The vacancy runs provide no $R_{\rm Q}$ controls. A dash marks an unavailable own-density control or derived comparison.}
\label{tab:cp2k-vacancies}
\small
\setlength{\tabcolsep}{4pt}
\begin{ruledtabular}
\begin{tabular}{lrrr rlr}
Configuration & $N_{\rm M}$ & $N_{\rm A}$ & $N_{\rm C}$ & $\eta_{N,\rm C}$ (\%) & First $10^{-5}$ crossing & $\Delta F/N_{\rm at}$ \\
\hline
B298, 1 vacancy, set 1 & \textbf{25} & 26 & 6 & 5.00 & 19/18/2 & $+0.023248$ \\
B298, 1 vacancy, set 2 & \textbf{24} & 26 & 6 & 10.00 & 17/17/2 & $+0.100059$ \\
B298, 1 vacancy, set 3 & \textbf{26} & 27 & 8 & 5.26 & 20/19/2 & $+0.067685$ \\
B298, 4 vacancies, set 1 & \textbf{27} & 29 & -- & -- & 20/19/-- & $+0.048199$ \\
B298, 4 vacancies, set 2 & \textbf{31} & \textbf{31} & -- & -- & 21/21/-- & $-0.273209$ \\
B298, 4 vacancies, set 3 & 28 & \textbf{27} & -- & -- & 20/20/-- & $-0.137410$ \\
\end{tabular}
\end{ruledtabular}
\end{table*}

\clearpage
\bibliography{references_qe_evidence_revision}

\end{document}